\pdfoutput=1
\documentclass[11pt,twoside,a4paper,pdftex,cmspaper,final,collab]{cms-tdr}

\begin{document}\cmsNoteHeader{CFT-09-018}
%
%
%

%
%
\hyphenation{env-iron-men-tal}
\hyphenation{had-ron-i-za-tion}
\hyphenation{cal-or-i-me-ter}
\hyphenation{de-vices}
%
%
\RCS$Revision: 1.30 $
\RCS$Date: 2010/01/12 00:25:27 $
\RCS$Name:  $
%
%
%

\newcommand {\etal}{\mbox{et al.}\xspace} 
\newcommand {\ie}{\mbox{i.e.}\xspace}     
\newcommand {\eg}{\mbox{e.g.}\xspace}     
\newcommand {\etc}{\mbox{etc.}\xspace}     
\newcommand {\vs}{\mbox{\sl vs.}\xspace}      
\newcommand {\mdash}{\ensuremath{\mathrm{-}}} 

\newcommand {\Lone}{Level-1\xspace} 
\newcommand {\Ltwo}{Level-2\xspace}
\newcommand {\Lthree}{Level-3\xspace}

\providecommand{\ACERMC} {\textsc{AcerMC}\xspace}
\providecommand{\ALPGEN} {{\textsc{alpgen}}\xspace}
\providecommand{\CHARYBDIS} {{\textsc{charybdis}}\xspace}
\providecommand{\CMKIN} {\textsc{cmkin}\xspace}
\providecommand{\CMSIM} {{\textsc{cmsim}}\xspace}
\providecommand{\CMSSW} {{\textsc{cmssw}}\xspace}
\providecommand{\COBRA} {{\textsc{cobra}}\xspace}
\providecommand{\COCOA} {{\textsc{cocoa}}\xspace}
\providecommand{\COMPHEP} {\textsc{CompHEP}\xspace}
\providecommand{\EVTGEN} {{\textsc{evtgen}}\xspace}
\providecommand{\FAMOS} {{\textsc{famos}}\xspace}
\providecommand{\GARCON} {\textsc{garcon}\xspace}
\providecommand{\GARFIELD} {{\textsc{garfield}}\xspace}
\providecommand{\GEANE} {{\textsc{geane}}\xspace}
\providecommand{\GEANTfour} {{\textsc{geant4}}\xspace}
\providecommand{\GEANTthree} {{\textsc{geant3}}\xspace}
\providecommand{\GEANT} {{\textsc{geant}}\xspace}
\providecommand{\HDECAY} {\textsc{hdecay}\xspace}
\providecommand{\HERWIG} {{\textsc{herwig}}\xspace}
\providecommand{\HIGLU} {{\textsc{higlu}}\xspace}
\providecommand{\HIJING} {{\textsc{hijing}}\xspace}
\providecommand{\IGUANA} {\textsc{iguana}\xspace}
\providecommand{\ISAJET} {{\textsc{isajet}}\xspace}
\providecommand{\ISAPYTHIA} {{\textsc{isapythia}}\xspace}
\providecommand{\ISASUGRA} {{\textsc{isasugra}}\xspace}
\providecommand{\ISASUSY} {{\textsc{isasusy}}\xspace}
\providecommand{\ISAWIG} {{\textsc{isawig}}\xspace}
\providecommand{\MADGRAPH} {\textsc{MadGraph}\xspace}
\providecommand{\MCATNLO} {\textsc{mc@nlo}\xspace}
\providecommand{\MCFM} {\textsc{mcfm}\xspace}
\providecommand{\MILLEPEDE} {{\textsc{millepede}}\xspace}
\providecommand{\ORCA} {{\textsc{orca}}\xspace}
\providecommand{\OSCAR} {{\textsc{oscar}}\xspace}
\providecommand{\PHOTOS} {\textsc{photos}\xspace}
\providecommand{\PROSPINO} {\textsc{prospino}\xspace}
\providecommand{\PYTHIA} {{\textsc{pythia}}\xspace}
\providecommand{\SHERPA} {{\textsc{sherpa}}\xspace}
\providecommand{\TAUOLA} {\textsc{tauola}\xspace}
\providecommand{\TOPREX} {\textsc{TopReX}\xspace}
\providecommand{\XDAQ} {{\textsc{xdaq}}\xspace}

\newcommand {\DZERO}{D\O\xspace}     


\newcommand{\de}{\ensuremath{^\circ}}
\newcommand{\ten}[1]{\ensuremath{\times \text{10}^\text{#1}}}
\newcommand{\unit}[1]{\ensuremath{\text{\,#1}}\xspace}
\newcommand{\mum}{\ensuremath{\,\mu\text{m}}\xspace}
\newcommand{\micron}{\ensuremath{\,\mu\text{m}}\xspace}
\newcommand{\cm}{\ensuremath{\,\text{cm}}\xspace}
\newcommand{\mm}{\ensuremath{\,\text{mm}}\xspace}
\newcommand{\mus}{\ensuremath{\,\mu\text{s}}\xspace}
\newcommand{\keV}{\ensuremath{\,\text{ke\hspace{-.08em}V}}\xspace}
\newcommand{\MeV}{\ensuremath{\,\text{Me\hspace{-.08em}V}}\xspace}
\newcommand{\GeV}{\ensuremath{\,\text{Ge\hspace{-.08em}V}}\xspace}
\newcommand{\TeV}{\ensuremath{\,\text{Te\hspace{-.08em}V}}\xspace}
\newcommand{\PeV}{\ensuremath{\,\text{Pe\hspace{-.08em}V}}\xspace}
\newcommand{\keVc}{\ensuremath{{\,\text{ke\hspace{-.08em}V\hspace{-0.16em}/\hspace{-0.08em}c}}}\xspace}
\newcommand{\MeVc}{\ensuremath{{\,\text{Me\hspace{-.08em}V\hspace{-0.16em}/\hspace{-0.08em}c}}}\xspace}
\newcommand{\GeVc}{\ensuremath{{\,\text{Ge\hspace{-.08em}V\hspace{-0.16em}/\hspace{-0.08em}c}}}\xspace}
\newcommand{\TeVc}{\ensuremath{{\,\text{Te\hspace{-.08em}V\hspace{-0.16em}/\hspace{-0.08em}c}}}\xspace}
\newcommand{\keVcc}{\ensuremath{{\,\text{ke\hspace{-.08em}V\hspace{-0.16em}/\hspace{-0.08em}c}^\text{2}}}\xspace}
\newcommand{\MeVcc}{\ensuremath{{\,\text{Me\hspace{-.08em}V\hspace{-0.16em}/\hspace{-0.08em}c}^\text{2}}}\xspace}
\newcommand{\GeVcc}{\ensuremath{{\,\text{Ge\hspace{-.08em}V\hspace{-0.16em}/\hspace{-0.08em}c}^\text{2}}}\xspace}
\newcommand{\TeVcc}{\ensuremath{{\,\text{Te\hspace{-.08em}V\hspace{-0.16em}/\hspace{-0.08em}c}^\text{2}}}\xspace}

\newcommand{\pbinv} {\mbox{\ensuremath{\,\text{pb}^\text{$-$1}}}\xspace}
\newcommand{\fbinv} {\mbox{\ensuremath{\,\text{fb}^\text{$-$1}}}\xspace}
\newcommand{\nbinv} {\mbox{\ensuremath{\,\text{nb}^\text{$-$1}}}\xspace}
\newcommand{\percms}{\ensuremath{\,\text{cm}^\text{$-$2}\,\text{s}^\text{$-$1}}\xspace}
\newcommand{\lumi}{\ensuremath{\mathcal{L}}\xspace}
\newcommand{\Lumi}{\ensuremath{\mathcal{L}}\xspace}
%
\newcommand{\LvLow}  {\ensuremath{\mathcal{L}=\text{10}^\text{32}\,\text{cm}^\text{$-$2}\,\text{s}^\text{$-$1}}\xspace}
\newcommand{\LLow}   {\ensuremath{\mathcal{L}=\text{10}^\text{33}\,\text{cm}^\text{$-$2}\,\text{s}^\text{$-$1}}\xspace}
\newcommand{\lowlumi}{\ensuremath{\mathcal{L}=\text{2}\times \text{10}^\text{33}\,\text{cm}^\text{$-$2}\,\text{s}^\text{$-$1}}\xspace}
\newcommand{\LMed}   {\ensuremath{\mathcal{L}=\text{2}\times \text{10}^\text{33}\,\text{cm}^\text{$-$2}\,\text{s}^\text{$-$1}}\xspace}
\newcommand{\LHigh}  {\ensuremath{\mathcal{L}=\text{10}^\text{34}\,\text{cm}^\text{$-$2}\,\text{s}^\text{$-$1}}\xspace}
\newcommand{\hilumi} {\ensuremath{\mathcal{L}=\text{10}^\text{34}\,\text{cm}^\text{$-$2}\,\text{s}^\text{$-$1}}\xspace}


\newcommand{\zp}{\ensuremath{\mathrm{Z}^\prime}\xspace}


\newcommand{\kt}{\ensuremath{k_{\mathrm{T}}}\xspace}
\newcommand{\BC}{\ensuremath{{B_{\mathrm{c}}}}\xspace}
\newcommand{\bbarc}{\ensuremath{{\overline{b}c}}\xspace}
\newcommand{\bbbar}{\ensuremath{{b\overline{b}}}\xspace}
\newcommand{\ccbar}{\ensuremath{{c\overline{c}}}\xspace}
\newcommand{\JPsi}{\ensuremath{{J}/\psi}\xspace}
\newcommand{\bspsiphi}{\ensuremath{B_s \to \JPsi\, \phi}\xspace}
\newcommand{\AFB}{\ensuremath{A_\mathrm{FB}}\xspace}
\newcommand{\EE}{\ensuremath{e^+e^-}\xspace}
\newcommand{\MM}{\ensuremath{\mu^+\mu^-}\xspace}
\newcommand{\TT}{\ensuremath{\tau^+\tau^-}\xspace}
\newcommand{\wangle}{\ensuremath{\sin^{2}\theta_{\mathrm{eff}}^\mathrm{lept}(M^2_\mathrm{Z})}\xspace}
\newcommand{\ttbar}{\ensuremath{{t\overline{t}}}\xspace}
\newcommand{\stat}{\ensuremath{\,\text{(stat.)}}\xspace}
\newcommand{\syst}{\ensuremath{\,\text{(syst.)}}\xspace}

\newcommand{\HGG}{\ensuremath{\mathrm{H}\to\gamma\gamma}}
\newcommand{\gev}{\GeV}
\newcommand{\GAMJET}{\ensuremath{\gamma + \mathrm{jet}}}
\newcommand{\PPTOJETS}{\ensuremath{\mathrm{pp}\to\mathrm{jets}}}
\newcommand{\PPTOGG}{\ensuremath{\mathrm{pp}\to\gamma\gamma}}
\newcommand{\PPTOGAMJET}{\ensuremath{\mathrm{pp}\to\gamma +
\mathrm{jet}
}}
\newcommand{\MH}{\ensuremath{\mathrm{M_{\mathrm{H}}}}}
\newcommand{\RNINE}{\ensuremath{\mathrm{R}_\mathrm{9}}}
\newcommand{\DR}{\ensuremath{\Delta\mathrm{R}}}


\newcommand{\PT}{\ensuremath{p_{\mathrm{T}}}\xspace}
\newcommand{\pt}{\ensuremath{p_{\mathrm{T}}}\xspace}
\newcommand{\ET}{\ensuremath{E_{\mathrm{T}}}\xspace}
\newcommand{\HT}{\ensuremath{H_{\mathrm{T}}}\xspace}
\newcommand{\et}{\ensuremath{E_{\mathrm{T}}}\xspace}
\newcommand{\Em}{\ensuremath{E\!\!\!/}\xspace}
\newcommand{\Pm}{\ensuremath{p\!\!\!/}\xspace}
\newcommand{\PTm}{\ensuremath{{p\!\!\!/}_{\mathrm{T}}}\xspace}
\newcommand{\ETm}{\ensuremath{E_{\mathrm{T}}^{\mathrm{miss}}}\xspace}
\newcommand{\MET}{\ensuremath{E_{\mathrm{T}}^{\mathrm{miss}}}\xspace}
\newcommand{\ETmiss}{\ensuremath{E_{\mathrm{T}}^{\mathrm{miss}}}\xspace}
\newcommand{\VEtmiss}{\ensuremath{{\vec E}_{\mathrm{T}}^{\mathrm{miss}}}\xspace}

%

\newcommand{\ga}{\ensuremath{\gtrsim}}
\newcommand{\la}{\ensuremath{\lesssim}}
\newcommand{\swsq}{\ensuremath{\sin^2\theta_W}\xspace}
\newcommand{\cwsq}{\ensuremath{\cos^2\theta_W}\xspace}
\newcommand{\tanb}{\ensuremath{\tan\beta}\xspace}
\newcommand{\tanbsq}{\ensuremath{\tan^{2}\beta}\xspace}
\newcommand{\sidb}{\ensuremath{\sin 2\beta}\xspace}
\newcommand{\alpS}{\ensuremath{\alpha_S}\xspace}
\newcommand{\alpt}{\ensuremath{\tilde{\alpha}}\xspace}

\newcommand{\QL}{\ensuremath{Q_L}\xspace}
\newcommand{\sQ}{\ensuremath{\tilde{Q}}\xspace}
\newcommand{\sQL}{\ensuremath{\tilde{Q}_L}\xspace}
\newcommand{\ULC}{\ensuremath{U_L^C}\xspace}
\newcommand{\sUC}{\ensuremath{\tilde{U}^C}\xspace}
\newcommand{\sULC}{\ensuremath{\tilde{U}_L^C}\xspace}
\newcommand{\DLC}{\ensuremath{D_L^C}\xspace}
\newcommand{\sDC}{\ensuremath{\tilde{D}^C}\xspace}
\newcommand{\sDLC}{\ensuremath{\tilde{D}_L^C}\xspace}
\newcommand{\LL}{\ensuremath{L_L}\xspace}
\newcommand{\sL}{\ensuremath{\tilde{L}}\xspace}
\newcommand{\sLL}{\ensuremath{\tilde{L}_L}\xspace}
\newcommand{\ELC}{\ensuremath{E_L^C}\xspace}
\newcommand{\sEC}{\ensuremath{\tilde{E}^C}\xspace}
\newcommand{\sELC}{\ensuremath{\tilde{E}_L^C}\xspace}
\newcommand{\sEL}{\ensuremath{\tilde{E}_L}\xspace}
\newcommand{\sER}{\ensuremath{\tilde{E}_R}\xspace}
\newcommand{\sFer}{\ensuremath{\tilde{f}}\xspace}
\newcommand{\sQua}{\ensuremath{\tilde{q}}\xspace}
\newcommand{\sUp}{\ensuremath{\tilde{u}}\xspace}
\newcommand{\suL}{\ensuremath{\tilde{u}_L}\xspace}
\newcommand{\suR}{\ensuremath{\tilde{u}_R}\xspace}
\newcommand{\sDw}{\ensuremath{\tilde{d}}\xspace}
\newcommand{\sdL}{\ensuremath{\tilde{d}_L}\xspace}
\newcommand{\sdR}{\ensuremath{\tilde{d}_R}\xspace}
\newcommand{\sTop}{\ensuremath{\tilde{t}}\xspace}
\newcommand{\stL}{\ensuremath{\tilde{t}_L}\xspace}
\newcommand{\stR}{\ensuremath{\tilde{t}_R}\xspace}
\newcommand{\stone}{\ensuremath{\tilde{t}_1}\xspace}
\newcommand{\sttwo}{\ensuremath{\tilde{t}_2}\xspace}
\newcommand{\sBot}{\ensuremath{\tilde{b}}\xspace}
\newcommand{\sbL}{\ensuremath{\tilde{b}_L}\xspace}
\newcommand{\sbR}{\ensuremath{\tilde{b}_R}\xspace}
\newcommand{\sbone}{\ensuremath{\tilde{b}_1}\xspace}
\newcommand{\sbtwo}{\ensuremath{\tilde{b}_2}\xspace}
\newcommand{\sLep}{\ensuremath{\tilde{l}}\xspace}
\newcommand{\sLepC}{\ensuremath{\tilde{l}^C}\xspace}
\newcommand{\sEl}{\ensuremath{\tilde{e}}\xspace}
\newcommand{\sElC}{\ensuremath{\tilde{e}^C}\xspace}
\newcommand{\seL}{\ensuremath{\tilde{e}_L}\xspace}
\newcommand{\seR}{\ensuremath{\tilde{e}_R}\xspace}
\newcommand{\snL}{\ensuremath{\tilde{\nu}_L}\xspace}
\newcommand{\sMu}{\ensuremath{\tilde{\mu}}\xspace}
\newcommand{\sNu}{\ensuremath{\tilde{\nu}}\xspace}
\newcommand{\sTau}{\ensuremath{\tilde{\tau}}\xspace}
\newcommand{\Glu}{\ensuremath{g}\xspace}
\newcommand{\sGlu}{\ensuremath{\tilde{g}}\xspace}
\newcommand{\Wpm}{\ensuremath{W^{\pm}}\xspace}
\newcommand{\sWpm}{\ensuremath{\tilde{W}^{\pm}}\xspace}
\newcommand{\Wz}{\ensuremath{W^{0}}\xspace}
\newcommand{\sWz}{\ensuremath{\tilde{W}^{0}}\xspace}
\newcommand{\sWino}{\ensuremath{\tilde{W}}\xspace}
\newcommand{\Bz}{\ensuremath{B^{0}}\xspace}
\newcommand{\sBz}{\ensuremath{\tilde{B}^{0}}\xspace}
\newcommand{\sBino}{\ensuremath{\tilde{B}}\xspace}
\newcommand{\Zz}{\ensuremath{Z^{0}}\xspace}
\newcommand{\sZino}{\ensuremath{\tilde{Z}^{0}}\xspace}
\newcommand{\sGam}{\ensuremath{\tilde{\gamma}}\xspace}
\newcommand{\chiz}{\ensuremath{\tilde{\chi}^{0}}\xspace}
\newcommand{\chip}{\ensuremath{\tilde{\chi}^{+}}\xspace}
\newcommand{\chim}{\ensuremath{\tilde{\chi}^{-}}\xspace}
\newcommand{\chipm}{\ensuremath{\tilde{\chi}^{\pm}}\xspace}
\newcommand{\Hone}{\ensuremath{H_{d}}\xspace}
\newcommand{\sHone}{\ensuremath{\tilde{H}_{d}}\xspace}
\newcommand{\Htwo}{\ensuremath{H_{u}}\xspace}
\newcommand{\sHtwo}{\ensuremath{\tilde{H}_{u}}\xspace}
\newcommand{\sHig}{\ensuremath{\tilde{H}}\xspace}
\newcommand{\sHa}{\ensuremath{\tilde{H}_{a}}\xspace}
\newcommand{\sHb}{\ensuremath{\tilde{H}_{b}}\xspace}
\newcommand{\sHpm}{\ensuremath{\tilde{H}^{\pm}}\xspace}
\newcommand{\hz}{\ensuremath{h^{0}}\xspace}
\newcommand{\Hz}{\ensuremath{H^{0}}\xspace}
\newcommand{\Az}{\ensuremath{A^{0}}\xspace}
\newcommand{\Hpm}{\ensuremath{H^{\pm}}\xspace}
\newcommand{\sGra}{\ensuremath{\tilde{G}}\xspace}
\newcommand{\mtil}{\ensuremath{\tilde{m}}\xspace}
\newcommand{\rpv}{\ensuremath{\rlap{\kern.2em/}R}\xspace}
\newcommand{\LLE}{\ensuremath{LL\bar{E}}\xspace}
\newcommand{\LQD}{\ensuremath{LQ\bar{D}}\xspace}
\newcommand{\UDD}{\ensuremath{\overline{UDD}}\xspace}
\newcommand{\Lam}{\ensuremath{\lambda}\xspace}
\newcommand{\Lamp}{\ensuremath{\lambda'}\xspace}
\newcommand{\Lampp}{\ensuremath{\lambda''}\xspace}
\newcommand{\spinbd}[2]{\ensuremath{\bar{#1}_{\dot{#2}}}\xspace}

\newcommand{\MD}{\ensuremath{{M_\mathrm{D}}}\xspace}
\newcommand{\Mpl}{\ensuremath{{M_\mathrm{Pl}}}\xspace}
\newcommand{\Rinv} {\ensuremath{{R}^{-1}}\xspace}

%
%
\hyphenation{en-viron-men-tal}

\cmsNoteHeader{09-018}
\title{Performance of CMS Hadron Calorimeter Timing and Synchronization using Test Beam, Cosmic Ray, and LHC Beam Data}

\address[cern]{CERN}
\address[nwu]{Northwestern University}
\address[umn]{University of Minnesota}
\address[fiu]{Florida International University}
\address[fit]{Florida Institute of Technology}
\address[fnal]{Fermi National Laboratory}
\author[fnal]{Jordan Damgov}\author[umn]{Phillip Dudero}\author[cern]{Richard Kellogg}\author[fiu]{Luis Lebolo}\author[umn]{Jeremiah Mans}\author[nwu]{Mayda Velasco}\author[fit]{Igor Vodopiyanov}%

\date{\today}

\abstract{This paper discusses the design and performance of the time
measurement technique and of the synchronization systems of the CMS
hadron calorimeter. Time measurement performance results are presented
from test beam data taken in the years 2004 and 2006. For hadronic
showers of energy greater than 100~GeV, the timing resolution is
measured to be about 1.2~ns. Time synchronization and out-of-time
background rejection results are presented from the Cosmic Run At Four
Tesla and LHC beam runs taken in the Autumn of 2008. The inter-channel
synchronization is measured to be within $\pm$2~ns.}

\hypersetup{%
pdfauthor={Phillip Dudero, Jeremiah Mans, Mayda Velasco},%
pdftitle={Performance of CMS Hadron Calorimeter Timing and Synchronization using Test Beam, Cosmic Ray, and LHC Beam Data},%
pdfsubject={CMS},%
pdfkeywords={Calorimeter, Calorimeter methods, Front-end electronics for detector readout}}

\maketitle 

\newenvironment{2figures}[1]{\begin{figure}[#1]
  \begin{center}
    \begin{tabular}{p{.47\textwidth}p{.47\textwidth}} }
 {  \end{tabular}
  \end{center}
 \end{figure}
}

\newenvironment{2figuresAdj}[3]{\begin{figure}[#1]
  \begin{center}
    \begin{tabular}{p{#2\textwidth}p{#3\textwidth}} }
 {  \end{tabular}
  \end{center}
 \end{figure}
}


\section{Introduction}

  The primary goal of the Compact Muon Solenoid (CMS)
  experiment \cite{cite:jinstcms} is to explore particle physics at
  the \TeV\ energy scale exploiting the proton-proton collisions
  delivered by the Large Hadron Collider
  (LHC) \cite{cite:jinstlhc}. The central feature of the CMS apparatus
  is a superconducting solenoid, of 6~m internal diameter, providing a
  field of 3.8~T.  Within the field volume are the silicon pixel and
  strip tracker, the crystal electromagnetic calorimeter (ECAL) and
  the brass/scintillator hadron calorimeter (HCAL). Muons are measured
  in gas-ionization detectors embedded in the steel return yoke.  In
  addition to the barrel and endcap detectors, CMS has extensive
  forward calorimetry.

  The primary purpose of the HCAL is the measurement of hadronic
  energy from collisions in CMS.  In addition to the energy
  measurement, the HCAL is also able to perform a precise time
  measurement for each energy deposit. Precise time measurements are
  valuable for excluding calorimeter noise and energy deposits from
  beam halo and cosmic ray muons.  Time information can also be valuable
  for identifying new physics signals such as long-lived particle decays
  and slow high-mass charged particles \cite{cite:cmsHSCPnote}.

  This paper is organized as follows. Section~\ref{sec:HCALdesc}
  reviews the pertinent details of the HCAL construction and their
  fundamental impact on timing resolution.  Section~\ref{sec:reco}
  discusses the method used to extract a time value from the digitized
  HCAL signal. In Section~\ref{sec:tbandsplash}, the validation of the
  method is presented based on measurements in test beam and initial
  beam operations of the LHC in September~2008.
  Section~\ref{sec:suppression} details the performance of timing
  filters in the suppression of non-collision-based backgrounds and
  the effect these timing filters have on simulated physics events.

\section{CMS Hadron Calorimeter Description}
\label{sec:HCALdesc}

A detailed description of the HCAL can be found elsewhere
\cite{cite:jinstcms}. Briefly, the HCAL consists of a set of sampling
calorimeters.  The barrel \cite{cite:epjcHB} and endcap
\cite{cite:jinstcmsHE} calorimeters utilize alternating layers of brass
as absorber and plastic scintillator as active material. The
scintillation light is converted by wavelength-shifting fibers
embedded in the scintillator and channeled to hybrid photodiode
detectors via clear fibers.  The outer calorimeter \cite{cite:epjcHO}
utilizes the CMS magnet coil/cryostat and the steel of the magnet
return yoke as its absorber, and uses the same active material and
readout system as the barrel and endcap calorimeters.  The forward
calorimeter is based on Cherenkov light production in quartz fibers
and is not discussed in this paper, due to its different signal time
structure.

\begin{figure}[htbp]
\begin{center}
    \resizebox{0.7\linewidth}{!}{\includegraphics{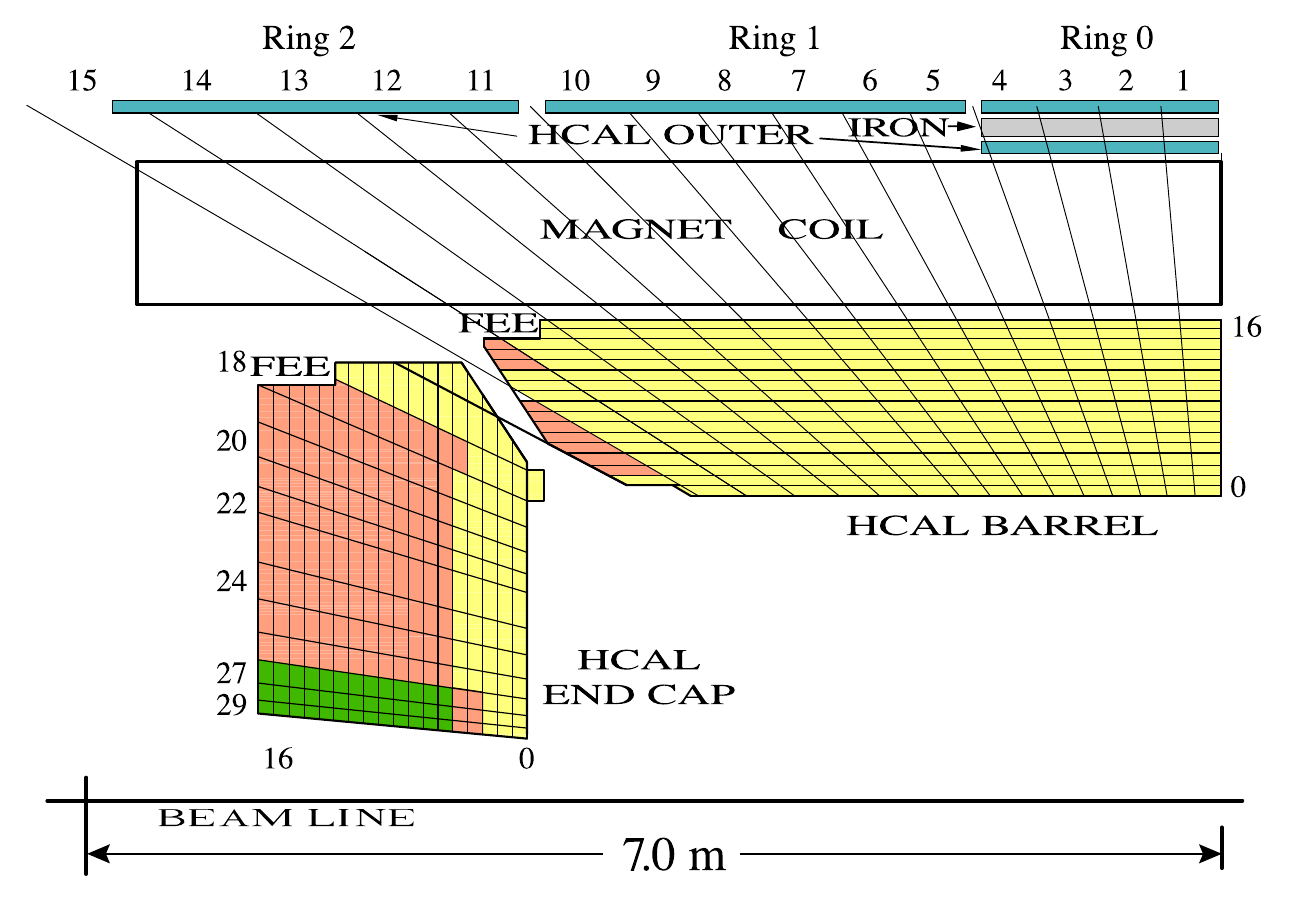}}

    \caption{A quarter slice of the CMS HCAL detectors. The right end
    of the beam line is the interaction point. FEE denotes the
    location of the Front End Electronics for the barrel and the
    endcap. In the diagram, the numbers on the top and left refer to
    segments in $\eta$, and the numbers on the right and the bottom
    refer to scintillator layers. Colors/shades indicate
    the combinations of layers that form the different depth
    segments, which are numbered sequentially starting at 1, moving
    outward from the interaction point. The outer calorimeter is
    assigned depth 4. Segmentation along $\phi$ is not shown.}

    \label{fig:HCALdiagram}
    \vspace{1cm}
    \resizebox{0.7\linewidth}{!}{\includegraphics{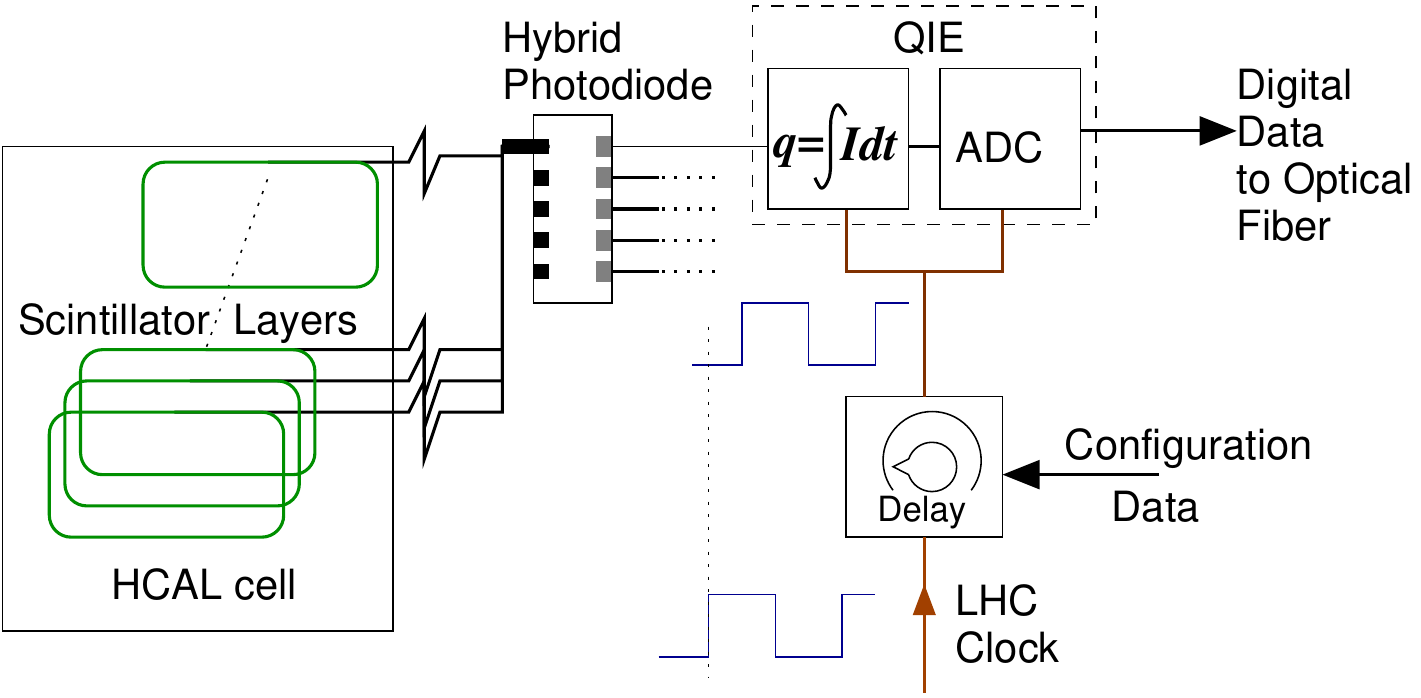}}

    \caption{A schematic view of the HCAL front-end readout electronics.
    The readout for one HCAL cell/channel is shown. Key features are
    the optical summing of layers, charge integration followed by
    sampling and digitization, and per-channel programmable delay
    settings. The ''QIE''~\cite{cite:qie} is a custom chip that
    contains the charge-integrating electronics with an analog-to-digital
    converter (ADC). The Configuration Data input defines the sampling
    delay settings.}

    \label{fig:HCALreadout}
\end{center}
\end{figure}

The HCAL is segmented into individual calorimeter cells along three
coordinates, $\eta$, $\phi$, and depth.  The $\phi$ coordinate is the
azimuthal angle and $\eta$ is the pseudorapidity.  The depth is an
integer coordinate that enumerates the segmentation longitudinally,
along the direction from the center of the nominal interaction
region. The layout of the barrel, endcap and outer calorimeter cells
is illustrated in Fig.~\ref{fig:HCALdiagram}.  Most cells include
several scintillator layers; for example, in most of the barrel all 17
scintillator layers are combined into a single depth segment.

Calorimetric measurements are acquired using the HCAL readout
electronics, shown schematically in Fig.~\ref{fig:HCALreadout}.
Each electronics channel collects and processes the signal of a single
cell. One calorimetric measurement is acquired with each LHC clock
tick (25~ns) from each cell in the HCAL. This defines a ``time
sample''. The HCAL front-end electronics does not sample the signal
instantaneously; rather, the electric current collected from the
photodetectors is integrated over each clock period and then sampled.
As a consequence, the sample clock is most commonly termed the
``integration clock''.

The integration clock can be delayed with respect to the LHC clock on
a per-channel basis by programmable settings, referred to as sampling
delay settings, that have a resolution of 1~ns.  The purpose of these
settings is to compensate for channel-dependent timing variations at
the hardware level, and to permit the energy estimate from each LHC
crossing to be reconstructed consistently for use in the Level-1
trigger for all pulse amplitudes and for all channels.  They also
provide an initial coarse timing calibration.


\subsection{Sources of timing variation and uncertainty}
\label{subsec:sources}

There are two dominant sources of channel-dependent timing variation:
the different time-of-flight of particles from the interaction region
to each HCAL cell and the different signal propagation times through
optical fibers of different lengths. Within the barrel, the first
effect tends to skew reconstructed times later for higher $\eta$. The
second effect, because of the location of the front-end electronics,
tends to skew reconstructed times earlier for higher $\eta$, and this
is the effect that dominates. The combination of these two effects
induces a timing dependence on $\eta$, with a spread in each
half-barrel of about 15~ns.  Smaller variations as a function of
$\phi$ are induced by clock distribution differences and other
effects.  By applying proper sampling delay settings, this spread can
be reduced substantially. In addition, the reconstruction software
utilizes a set of per-channel calibration constants to synchronize the
mean timing of energy measurements to less than 1~ns.

Fiber lengths also differ within each calorimeter cell along the
radial coordinate, but in this case there are no means available to
compensate for these differences. Since the signals from all the
scintillator layers comprising one cell are optically summed, and the
optical path lengths are not equalized across the layers, the
resulting signal is smeared in time. For hadrons, which exhibit large
shower-to-shower fluctuations in longitudinal development, the optical
summing of layers imposes a limit on the timing resolution, estimated
to be approximately 1~ns.  For signals with uniform energy deposit in
each layer (such as those arising from the beam-collimator
interactions described later), the resolution is not limited by
shower-development fluctuations and can be significantly smaller than
1~ns.

\section{HCAL Time Reconstruction}
\label{sec:reco}

When the Level-1 trigger system \cite{cite:craftL1trig} identifies an
event of potential physics interest, a set of 10 consecutive time
samples per channel containing the triggered bunch-crossing is
collected and sent to the high-level trigger software. The capability
of the HCAL system to reconstruct the arrival time of the signal more
precisely than the sample clock period derives from the spread of the
HCAL pulse shape over three to four time-integrated samples
(Fig.~\ref{fig:HCALpulse}). The time reconstruction software
calculates a first order time estimate from a center-of-gravity
technique using the three samples centered on the peak,

\begin{equation}\label{eq:wpkbin}
\textrm{Weighted peak bin} =
   \bigg[\frac{(p-1)A_{p-1} + pA_p + (p+1)A_{p+1}}{A_{p-1} + A_p + A_{p+1}}\bigg]\times C\quad,
\end{equation}

where $A_i$ represents the amplitude of an arbitrary time sample $i$,
and $p$ is the value of $i$ such that $A_p$ is maximum over the set of
samples.  In the case of multiple samples with the same amplitude, the
earliest one is picked. The constant $C$ is an amplitude-independent
normalization constant that rescales the first order estimate to a
range from zero to one.  The weighted peak bin is then used to determine
a second order correction (Fig.~\ref{fig:corfun}) that compensates
for the asymmetry of the pulse shape, yielding the phase of the signal
within the peak time sample.  An additional additive correction
determined in test bench measurements~\cite{cite:timeslew},
\begin{equation}
\Delta_\mathrm{slew} = (3.59\;\mathrm{ns}) \log_{10} \left[ \frac{1\TeV}{E (\TeV)} \right]\quad,
\end{equation}
compensates for an electronics effect that delays the measured time
for signals with pulse energy $E$ less than 1\TeV. The maximum value of
$\Delta_\mathrm{slew}$ is taken to be 10~ns.

This algorithm yields accurate time measurements except when
consecutive bunch crossings yield energy measurements of similar
amplitude within the same HCAL cell.  Such events can be identified by
their anomalous pulse shape, and the time measurement can be marked as
having poor resolution.

\begin{2figures}{htbp}
    \resizebox{\linewidth}{0.7\linewidth}{\includegraphics{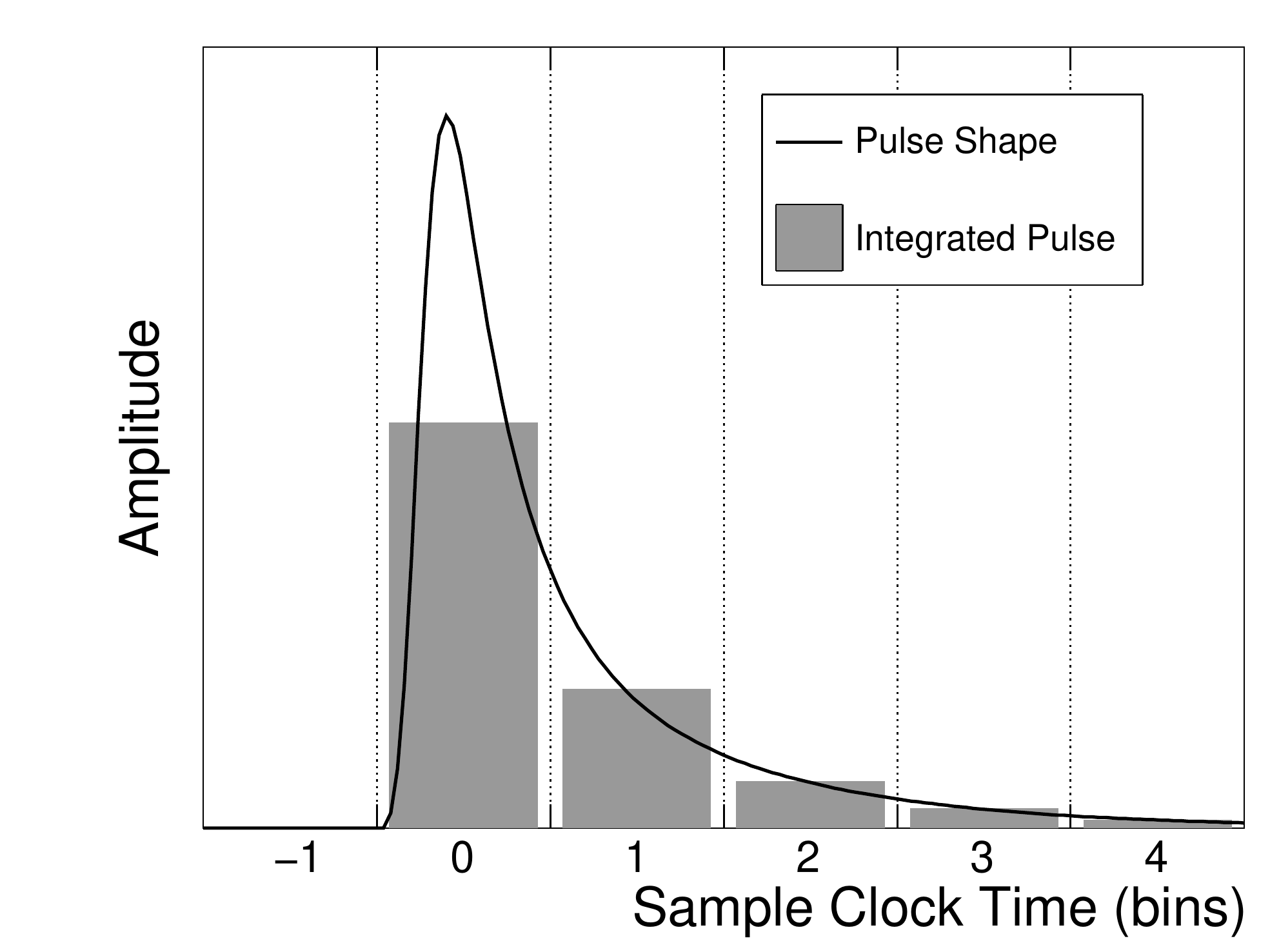}} &
    \resizebox{\linewidth}{0.7\linewidth}{\includegraphics{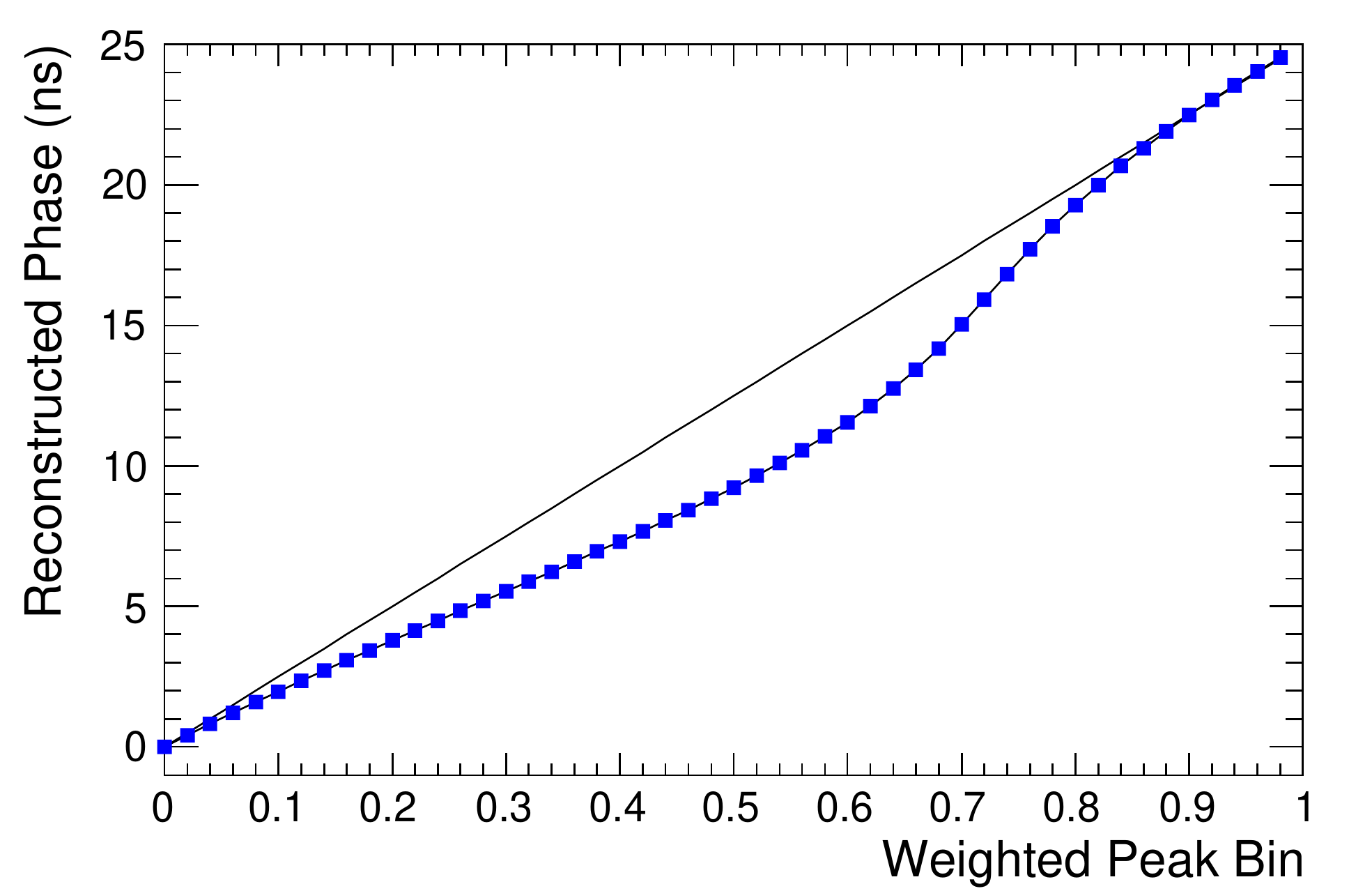}} \\
    \caption{The HCAL pulse shape and its relation to the integrated samples.
     Time sample 0 is defined by the trigger for a nominally synchronized event.}
    \label{fig:HCALpulse} &
    \caption{The correction function applied in signal reconstruction
    software to compensate for pulse shape asymmetry. The straight line
    represents a trivial correction function for a perfectly symmetric pulse.}
    \label{fig:corfun} \\
\end{2figures}

\section{Detector Timing/Synchronization Commissioning}
\label{sec:tbandsplash}

This section describes the performance characterization and
commissioning results for HCAL timing and synchronization. Beam tests
at the CERN SPS H2 area (hereafter referred to as ``test beam'')
produced key results \cite{cite:epjcHBTB} that define the timing
performance of the HCAL barrel and endcap; these results are presented
first. The results from the LHC 2008 beam commissioning data for the
barrel, endcap and outer calorimeters are then described; these results
validate and extend the initial synchronization calibration constants
determined from beam tests.

\subsection{H2 beam test data}

For test beams in 2004 and 2006, a section of the HCAL barrel was
mounted on a table that was adjustable in $\eta$ and $\phi$
coordinates. The unit under test was exposed to pion beams with
energies in the 3--300\GeV\ range.  The delivered beam was asynchronous
with the 40~MHz integration clock, so all relative sampling phases
could be investigated.  From these tests the following important
timing benchmarks for the HCAL were accurately measured.

\begin{2figures}{htbp}
    \resizebox{0.95\linewidth}{!}{\includegraphics{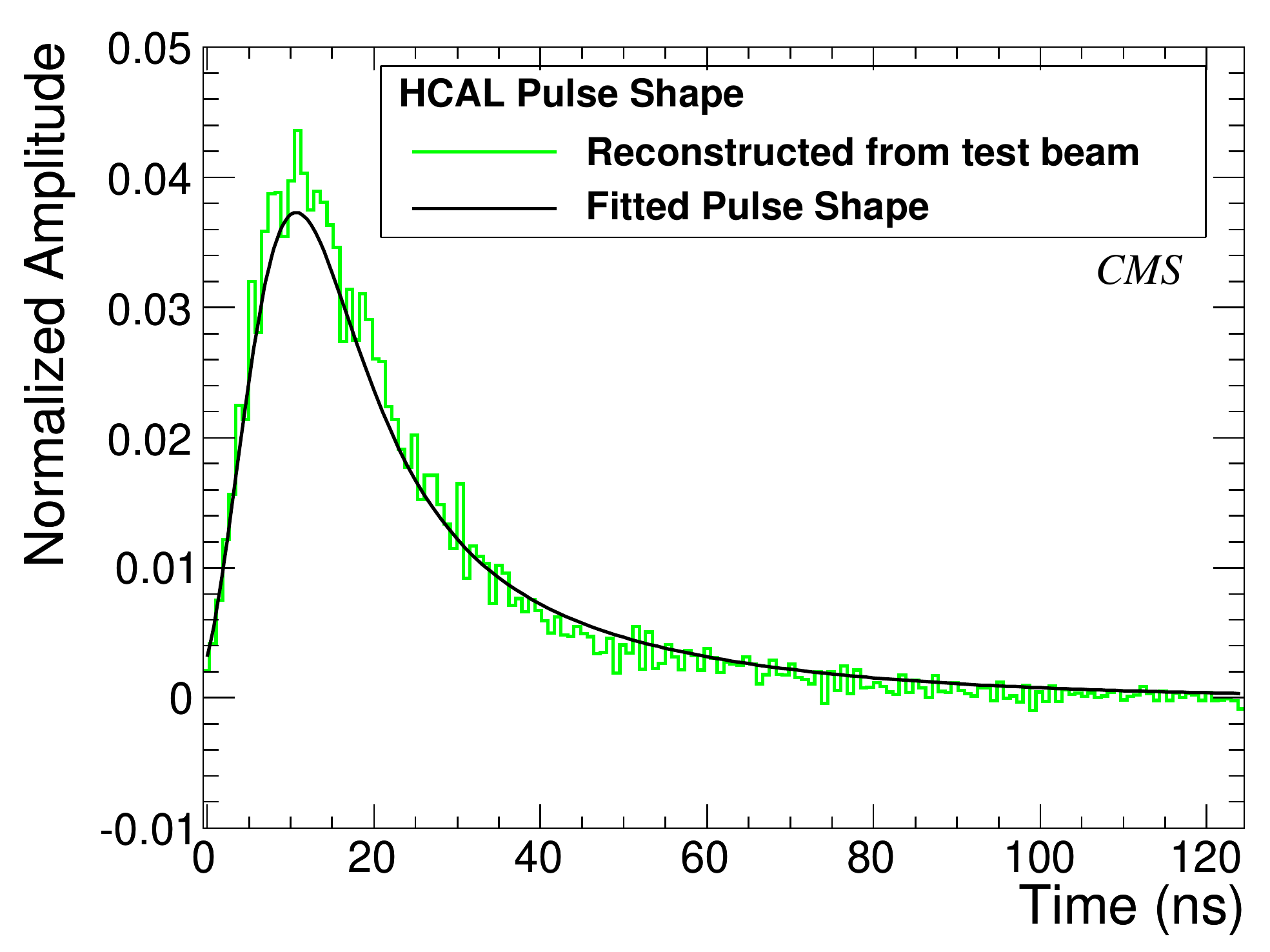}} &
    \resizebox{0.95\linewidth}{!}{\includegraphics{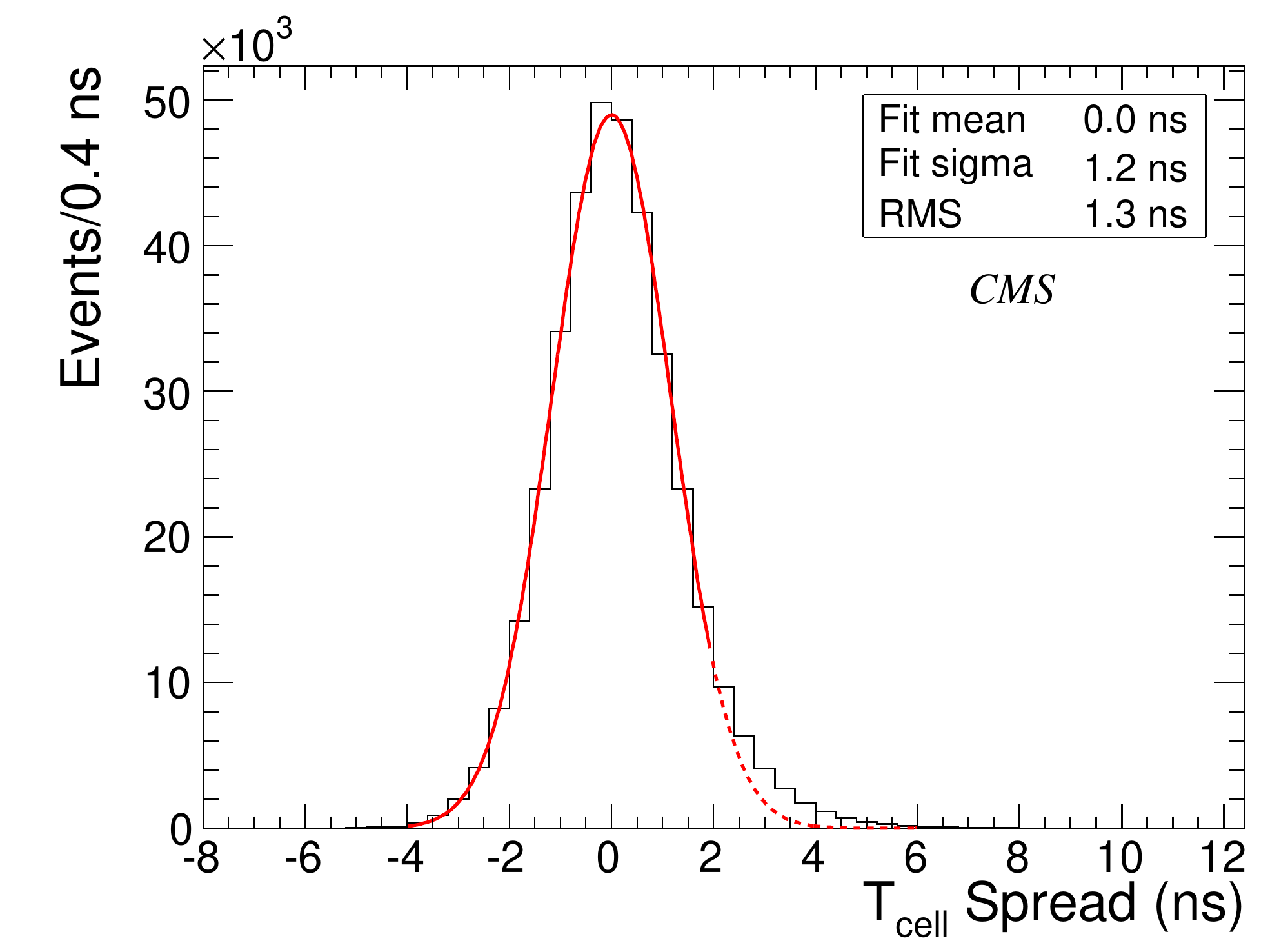}} \\
    \caption{Average pulse shape reconstructed for the HCAL from
             H2 beam test data using 300\GeV\ pions.}
    \label{fig:recopulses} &
    \caption{Resolution performance of the HCAL barrel from 150\GeV\
     pion test beam data.  The non-Gaussian tail is attributed to
     beam contamination from electrons and photons that shower
     in the first layers of the HCAL and thus reconstruct with later
     signal times.}
    \label{fig:Tspread} \\
\end{2figures}

The HCAL pulse shape was successfully reconstructed from data. Since
the pulse is integrated before it is digitized, this required a
numerical differentiation technique that utilized the phase between
particle arrival and the integration clock, as measured by a
time-to-digital converter (TDC) running at 32 times faster than the
sample clock rate. This allowed the separation of the data sample into
subsamples $S_n$, with~$n=$1..32 ordered by TDC phase. These
subsamples represent the integration of the pulse $P(t)$ shifted
successively by the TDC bit resolution ($\Delta t = 0.78$~ns).  Then,

\begin{equation}
\label{eq:recopulse}
\langle I_m \rangle_{S_{n+1}}-\langle I_m \rangle_{S_n} \simeq \int_\tau^{\tau+\Delta t}P(t)dt \simeq P_\mathrm{avg}(\tau)\Delta t\quad,
~~~~~~P_\mathrm{avg}(\tau)\simeq\frac{\Delta \langle I \rangle}{\Delta t}\bigg\vert_\tau\quad,
\end{equation}

where $\langle I_m \rangle_{S_n}$ is the average integrated amplitude
for time bin $m$ of data subsample $S_n$, $\tau=m\Delta t$ is the
phase within that time bin, and $P_\mathrm{avg}(\tau)$ is the average
value of $P(t)$ over $\Delta t$ at
$t=\tau$. Equation \ref{eq:recopulse} is valid to reconstruct the
first 25~ns of the pulse shape, for which $m=0$.  For each
remaining 25~ns interval, bin $m$ is incremented and a second integral
term appears in the equation that corresponds to the $\Delta t$-sized
portion of the pulse ``leaving'' the time bin. In this manner, the
whole pulse was reconstructed to approximately 0.8~ns resolution, as
is shown in Fig.~\ref{fig:recopulses}. The functional form matched to
these results is used in the simulation of the CMS detector response, and
also to determine the second-order time reconstruction correction
function (Fig.~\ref{fig:corfun}).

To measure both the channel-to-channel variations and the hadronic
timing resolution limit of the HCAL, a 150\GeV\ pion beam was directed
at every cell in the unit under test. From these data the variation of
the timing of signals as a function of $\eta$ between HCAL cells in
the barrel was mapped.  As the test setup mimicked the geometry and
fiber lengths of the final barrel detector, a set of final sampling
delay settings could be derived.  Section \ref{sec:splash} discusses
the performance of these settings with LHC beam data. The HCAL timing
resolution was determined by removing the channel-to-channel
variations and combining the data.  A Gaussian function was then
fitted to the distribution of the leading energy deposit from each
event, producing a best-fit width of $\sigma=1.2$~ns
(Fig.~\ref{fig:Tspread}).  This result is consistent with the expected
effect of the uncorrectable sources of timing spread described in
Sec.~\ref{subsec:sources}, averaged over many hadronic shower events.
The measured resolution is consistent across the full set of channels studied.

\begin{figure}[htbp]
  \centering
  \includegraphics[width=0.48\textwidth]{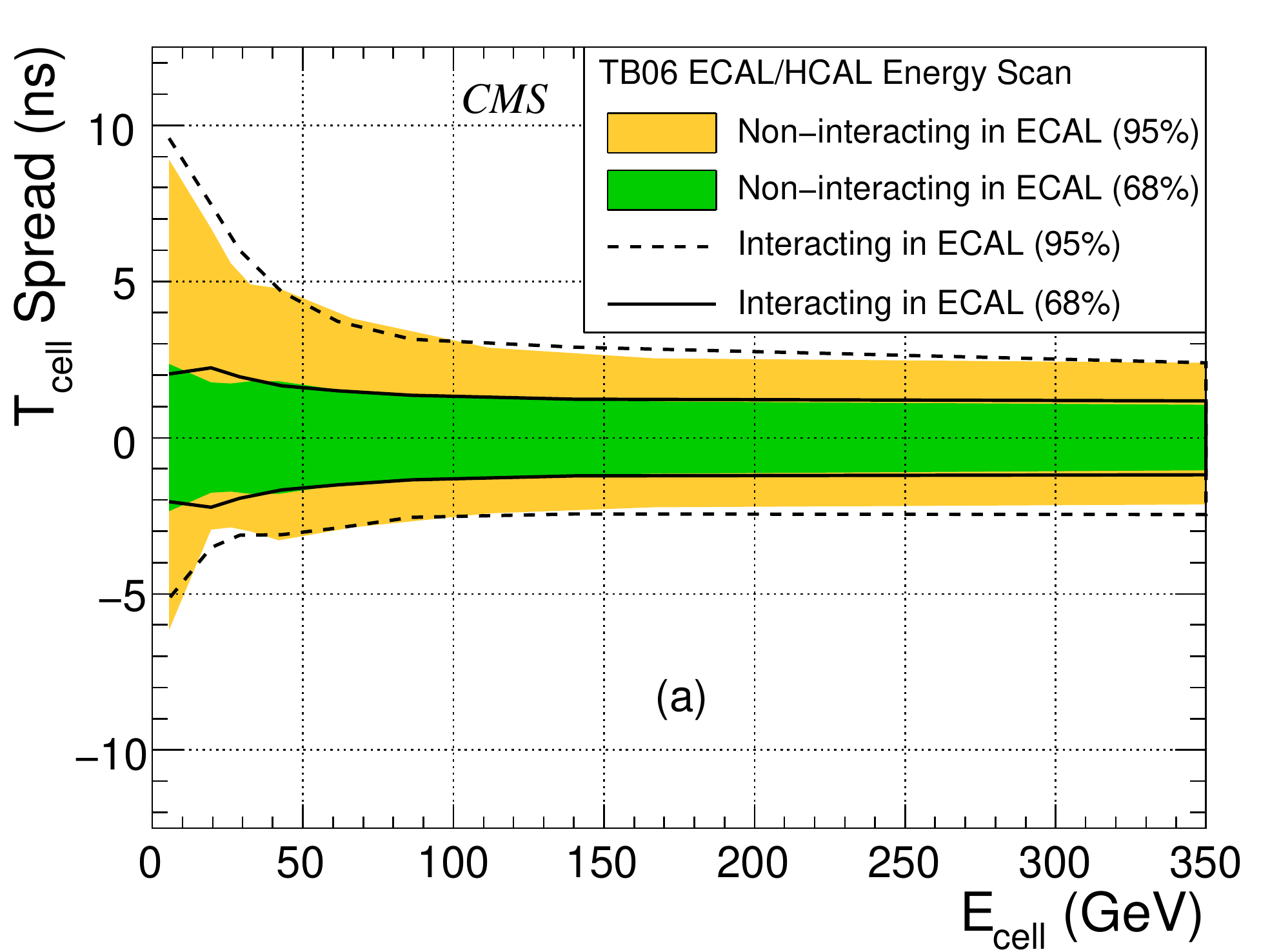}
  \hfill
  \includegraphics[width=0.48\textwidth]{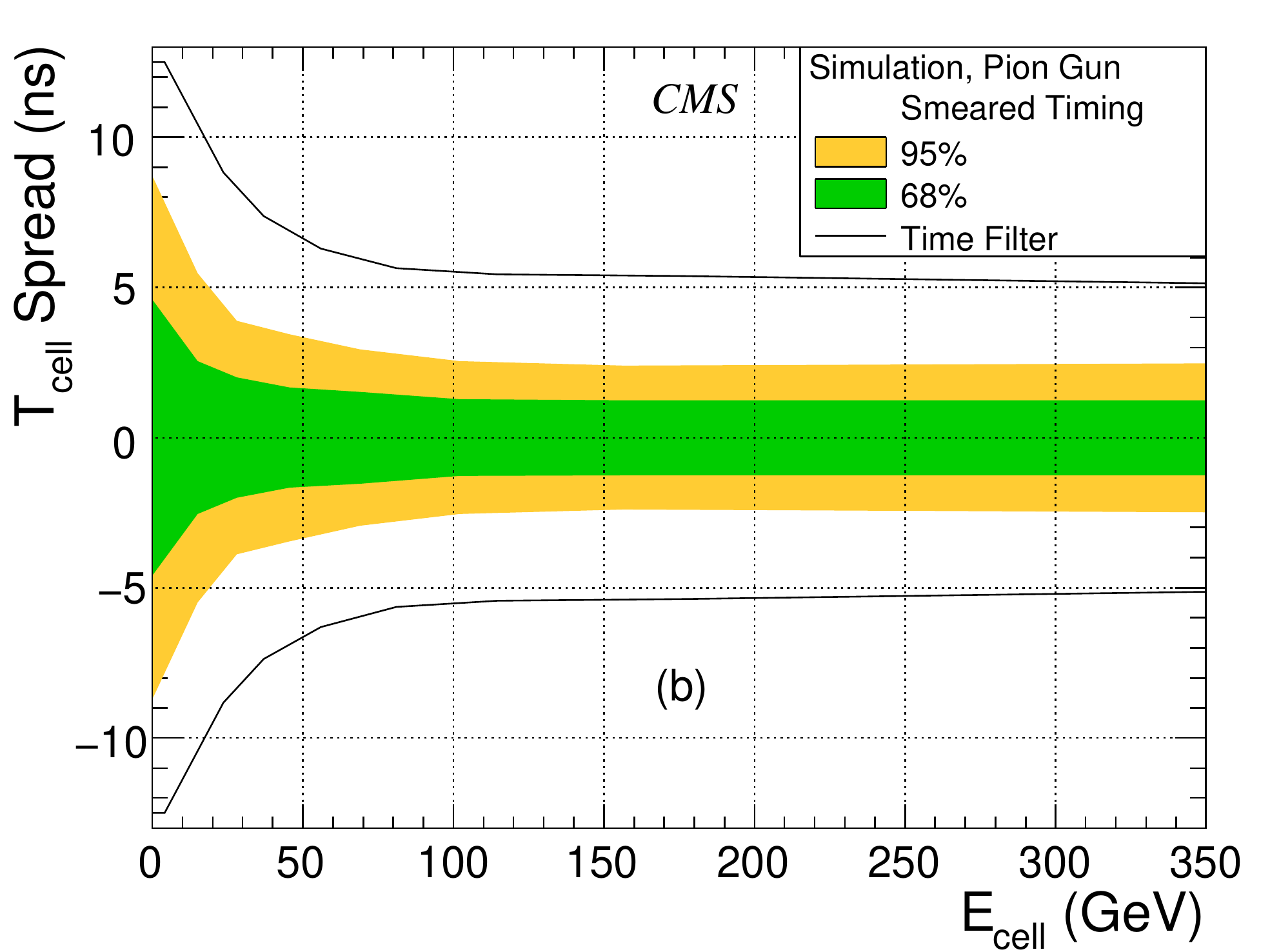}
  \caption{(a) Timing resolution as a function of energy measured
     during test beam runs in 2006, showing the consistency of time reconstruction
     for particles that begin showering in the ECAL (lines) and those that do not (areas).
     (b) Timing resolution as a function of energy of reconstructed deposits
     generated from CMS simulations. The lines marked ``Time Filter''
     indicate the boundaries of a timing window discussed later and used to accept or
     reject deposits for distinguishing signal from background.}
  \label{fig:timeResEnvelopesAndFilter}
\end{figure}

The variation of HCAL timing resolution as a function of energy was
also measured with pion beam data. In this measurement the HCAL unit under
test was paired with its corresponding section of the ECAL in front,
as in the CMS detector.  Events in two categories were considered:
those with no interaction in the ECAL ($<$1\GeV\ energy deposited) and
those with a significant interaction ($>$5\GeV\ energy deposited).  The
results indicate that the presence of the ECAL does not substantially
alter the timing resolution of the HCAL as a function of energy
(Fig.~\ref{fig:timeResEnvelopesAndFilter}(a)).  CMS software was subsequently adjusted by
smearing the times of simulated HCAL energy deposits so that simulated
data of the test beam configuration exhibited the same
energy-dependent time resolution (Fig.~\ref{fig:timeResEnvelopesAndFilter}(b)).

\subsection{LHC beam data}
\label{sec:splash}

Data from LHC beam commissioning were used to validate the barrel
sampling delay settings derived from test beam data, and to derive
settings to compensate for the $\eta$ dependence of timing in the
endcap and outer calorimeters. Data from LHC beam operations were
recorded for the first time on September 10, 2008. In preparation for
the arrival of LHC beam at CMS, the $\eta$-dependent sampling delays
measured from test beam were loaded into the barrel front-end
electronics. These delay values had been calculated to synchronize
channels for data from collisions. Delay values for endcap and outer
sectors were set to zero, since no timing measurements for these
sectors were available at the time.

\begin{figure}[htbp]
  \begin{center}
    \resizebox{\linewidth}{!}{\includegraphics{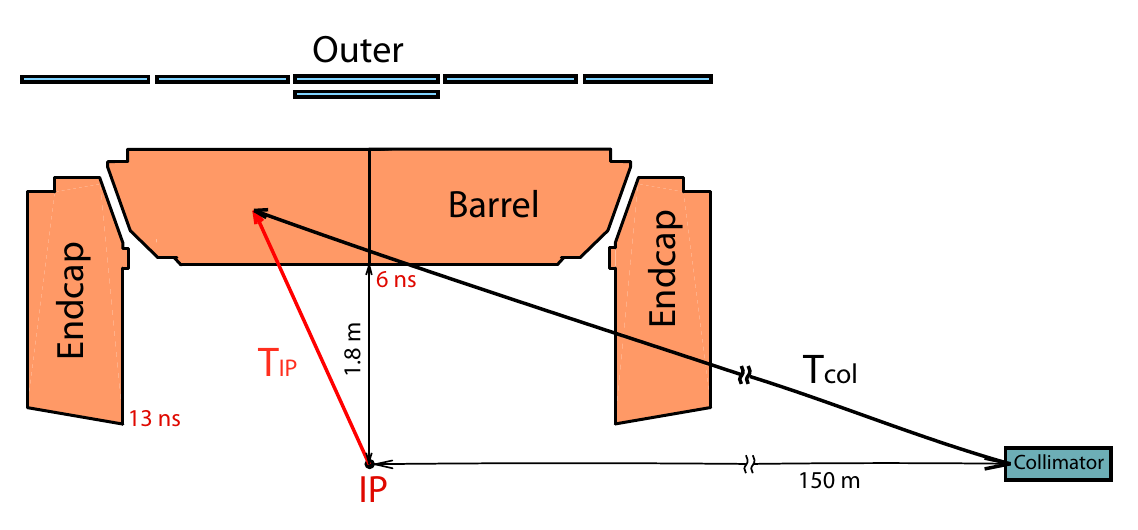}}

    \caption{A schematic view of the geometry of ``beam splash'' events.
     This geometry was used to predict the timing of
     energy deposits and thereby highlight channels in the HCAL that
     required synchronization.  The times $T_\mathrm{IP}$ and $T_\mathrm{col}$ refer to
     the times-of-flight of relativistic particles from the interaction
     point (IP) and collimator, respectively. The timing numbers give
     the specific values of $T_\mathrm{IP}$ to the front faces of the barrel
     and endcap.}

  \label{fig:JordansSplashDiag}
 \end{center}
\end{figure}

The LHC beam delivered to CMS consisted of a single 450\GeV\ proton
bunch circulating in the ring, first in one direction and then in the
other. The LHC beam was also steered into collimators located 150 m
upstream in either direction of the detector interaction point.  Such
events are referred to as ``beam splash'' events. Because of the large
amount of material in the collimator and along the path of the
secondaries, only muons were able to penetrate the entire CMS detector
and form the signal in the calorimeter system.  The millions of muons produced
from the dumping of the beam passed through the entire detector, depositing
large signals in every cell of the HCAL, with equivalent energy
ranging between hundreds of~\GeV\ to a few~\TeV.  A schematic
representation of the geometry of the ``beam splash'' setup is shown
in Fig.~\ref{fig:JordansSplashDiag}.

\begin{figure}[htbp]
  \centering
  \includegraphics[width=0.48\textwidth]{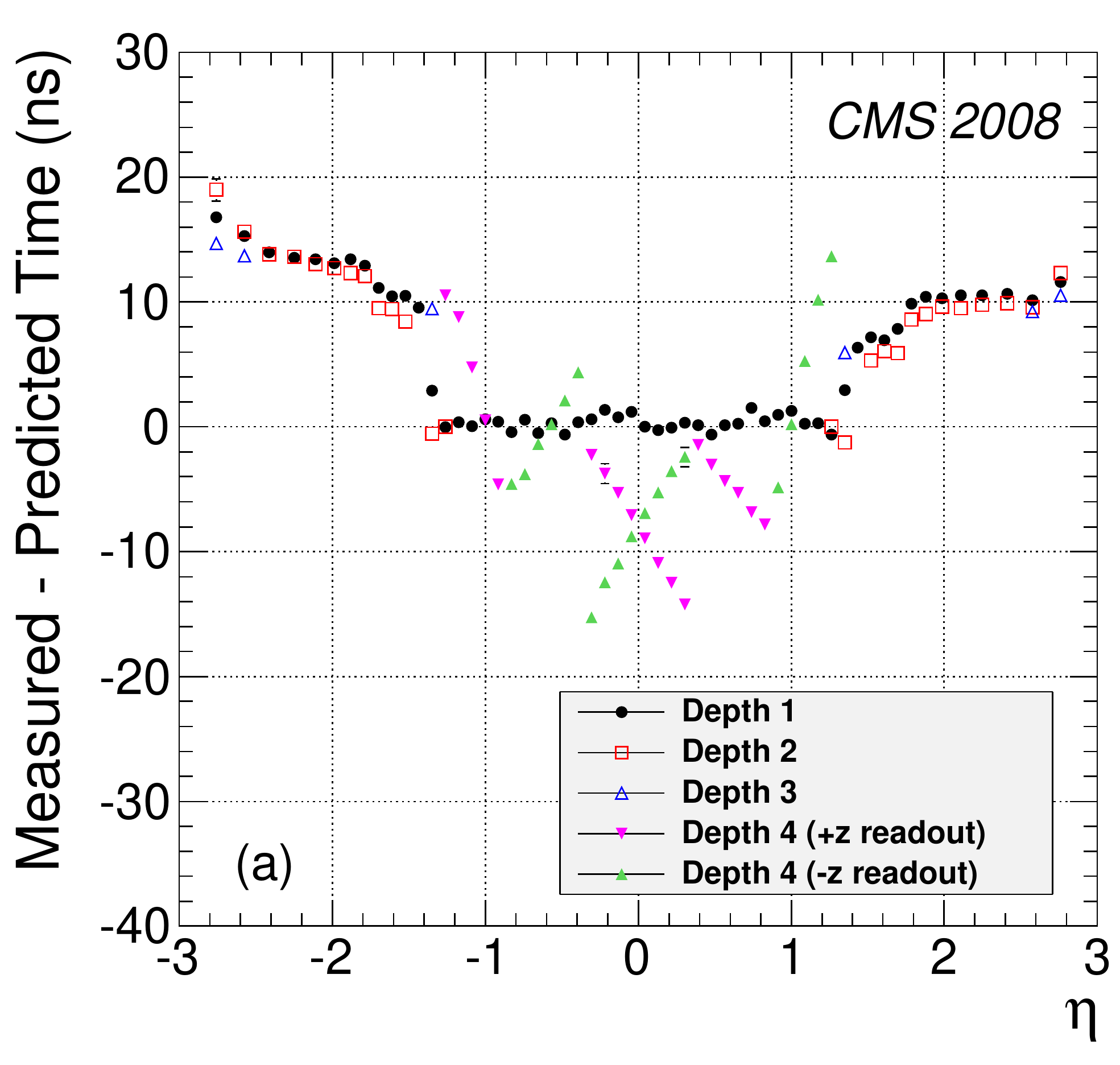}
  \hfill
  \includegraphics[width=0.48\textwidth]{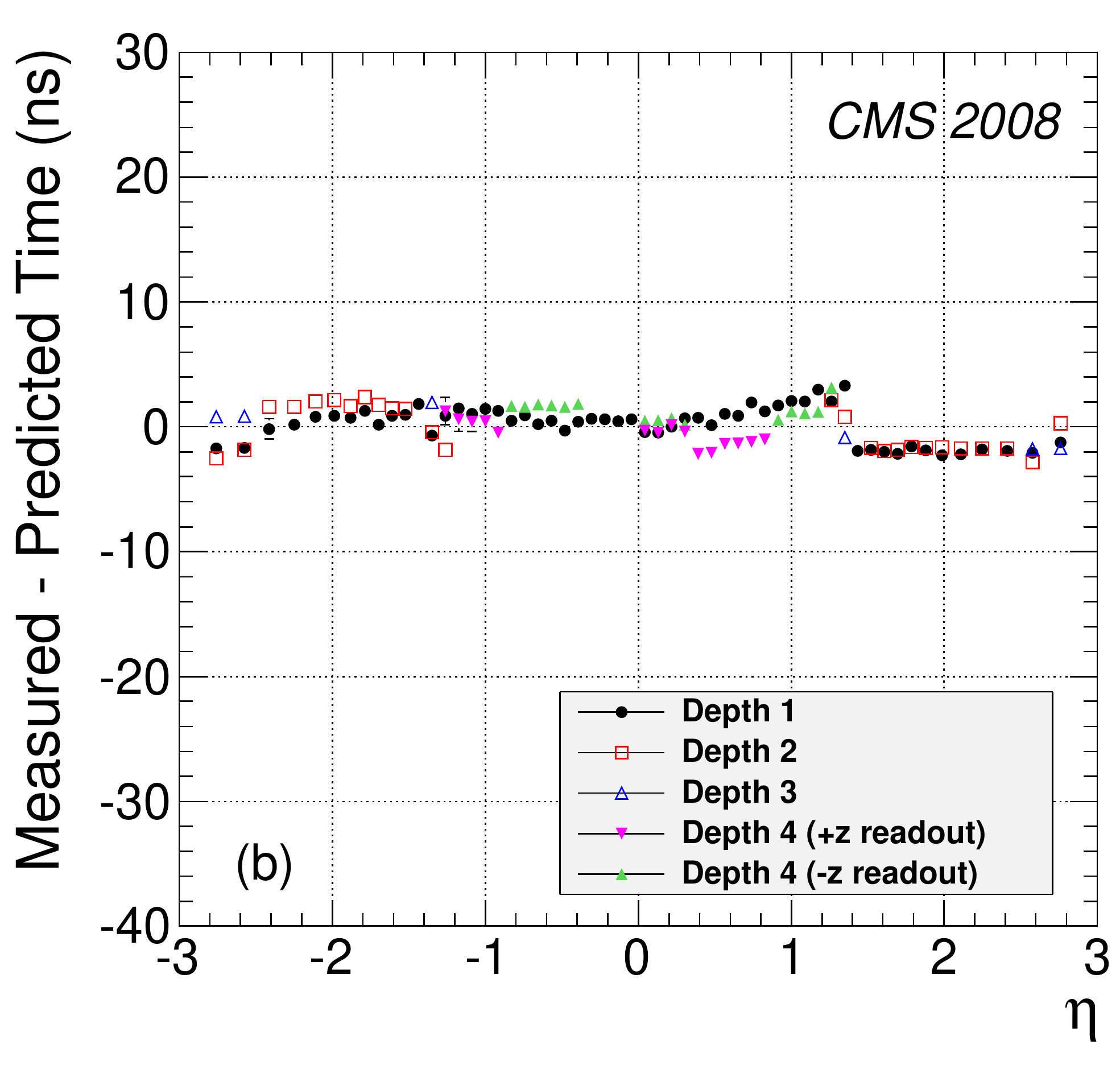}
  \caption{Difference between measured and predicted time as a function of $\eta$ from
  LHC ``beam splash'' events. (a) Per-channel delay settings that
  compensate for timing variations along $\eta$ were loaded into the
  front-end electronics for the barrel (depths 1--2, $|\eta|<1.4$)
  but not for the endcap (depths 1--3, $|\eta|>1.4$) or outer (depth 4)
  calorimeters. Events from both $+z$ and $-z$ directions are
  included.  (b) Compensating sample delay settings were loaded for
  all three calorimeters, barrel, endcap and outer. Only events
  from $+z$ direction were available and are included.}

 \label{fig:uncorrectedVsCorrectedSplash}
\end{figure}

The signal from ``beam splash'' muons has a well-defined time with
respect to the arrival of the LHC bunch and to the LHC clock.  The
time-of-flight of the muons from the collimators to the geometrical
centers of each cell was calculated and used to predict the expected
timing versus $\eta$ for all HCAL cells.  The data were averaged over
all $\phi$ values at each $\eta$ value to obtain an HCAL $\eta$ timing
profile.  The resulting distribution of the differences between the
measured and predicted times is shown in
Fig.~\ref{fig:uncorrectedVsCorrectedSplash}(a). The timing
measurements for the barrel region are distributed around zero within
$\pm1$~ns, demonstrating the correctness of the delay settings
calculated from beam tests. The timing for the endcap before any phase
corrections exhibits an average time difference that is 10~ns later
than the barrel. The x-shaped pattern and the structure at
$|\eta|>1.4$ in the outer calorimeter are understood to originate from
the pattern of optical fibers in the scintillator trays and the
positioning of readout electronics.

From these data, $\eta$-dependent sampling delays were derived for the
endcap and outer calorimeters.  The delays were loaded into the
front-end electronics, and new ``beam splash'' data were taken from
interactions on the $\emph{+z}$ collimator on September 18, 2008. The
same analysis used to derive the original delays was repeated on these
data; the results are shown in
Fig.~\ref{fig:uncorrectedVsCorrectedSplash}(b).  Some outer
calorimeter data points are omitted from the plot due to a technical
issue at the time of data-taking that has since been resolved.

Figure~\ref{fig:uncorrectedVsCorrectedSplash}(b) shows that, as a result
of the sampling delay derivation from ``beam splash'' data, barrel and
endcap channel synchronization along $\eta$ is verified to within
$\pm2$~ns.  Although ``beam splash'' event data from both directions
were used to derive the settings for endcap and outer, only event data
from the $\emph{+z}$ direction were available to verify these
settings. For this reason Fig.~\ref{fig:uncorrectedVsCorrectedSplash}(b)
exhibits systematic deviations, particularly near the barrel-endcap
boundary.  For cells that are significantly slanted relative to the
arrival direction of the muons, the ``beam splash'' will illuminate
the various layers of the cell at different times.  The effect is
canceled when considering events from both directions.  Additional
possible systematic effects, such as an offset in timing between the
positive and negative endcaps, will be studied with future ``beam
splash'' data from both directions.

The results shown in the last two sections indicate that the methods
of time reconstruction and synchronization are effective both in the
H2 test beam environment and with the HCAL integrated in CMS. The
final offline time corrections, which will be determined with
collision data when they become available, are expected to provide
synchronization with a spread in the per-channel mean times of less
than 1~ns.

\section{Impact of Timing in Analysis}
\label{sec:suppression}

Many interesting physics processes that the CMS experiment was
designed to study or discover contain the signature of an undetected
particle. Examples include the Standard Model neutrino and the
lightest stable particle in several hypothetical supersymmetric
particle spectra. Such particles would pass undetected through CMS,
carrying some of the energy and momentum that would otherwise
counterbalance the other collision products registered in the
calorimeters.  At hadron colliders, the initial energy and
longitudinal momentum of the colliding hadronic constituents are not
known, but the initial transverse momentum is known to be negligible
in comparison and can be treated as zero.  Therefore, the benchmark
used to infer the existence of an unobserved particle in any given
event is the calorimetric missing transverse
energy \cite{cite:cmsMETnote}.

The missing transverse energy (MET) is reconstructed from calorimeter
towers, which are reconstructed objects containing the sum of the
signals from HCAL and ECAL cells at the same $(\eta,\phi)$
coordinates. The MET is calculated by taking the magnitude of the
vector sum of the transverse energy contributions of towers,

\begin{equation}
\label{eq:met}
MET = \sqrt{(\sum_\mathrm{towers}E_i\sin\theta_i\cos\phi_i)^2 + (\sum_\mathrm{towers}E_i\sin\theta_i\sin\phi_i)^2}\quad,
\end{equation}

where $E_i$ and $(\theta_i,\phi_i)$ are the energy and angle coordinates
of each tower.

Multiple complications arise in the accurate measurement of MET.  The
energies of jets or leptons can be mismeasured due to detector
miscalibration or shower fluctuations in the dead material of the
detector. The MET in a collision event can also be affected by energy
deposits in HCAL cells from non-collision sources. Without a separate
identification of these sources the energy will be misassigned to the
event of interest, inducing fake MET. Example sources include cosmic
ray muons, detector noise, and beam halo.

This section explores the use of HCAL timing as a potential tool to
identify sources of fake MET. The CMS design ensures that the
phase of collision products relative to the LHC clock is fixed to high
precision.  The results presented in Section \ref{sec:tbandsplash}
indicate that the HCAL system is capable of being synchronized to
approximately 1--2~ns.  Since sources of fake MET are generally
uncorrelated with the collision event one can significantly suppress
their contributions to MET using their reconstructed times. In
the case of beam halo particles, although their arrival is correlated
to the arrival of proton bunches in the beam, the exact arrival time
to a given detector cell is generally distinguishable from the arrival
time of particles produced in the interaction region.

\subsection{Impact on out-of-time background}
\label{sec:bkgdtiming}

The strategy for using timing to reject backgrounds is implemented at
the level of reconstructed cell measurements, rather than higher-level
objects like hadronic jets or MET.  It is expected that during data
taking triggered events will contain collision data in around 1000
HCAL cells, over which additional out-of-time energy deposits may be
superimposed.  These out-of-time deposits can be individually removed
from the calculation of MET rather than rejecting the entire event.
The application of such a technique is appropriate for the analysis of
prompt physics signals, while an analysis focused on long-lived
particles might perform the reverse selection, preferring energy
deposits that are out of time.

The sources of irreducible timing uncertainty described in Section 2
will limit the tightness of the timing requirements that can be
applied. The degradation of time resolution for low energy particles
requires that any time filtering must take this energy dependence into
account in order to avoid biasing the measurement.  The timing filter
for individual cell measurements was defined according to the
resolution measurement shown in
Fig.~\ref{fig:timeResEnvelopesAndFilter}. The filter, shown in
Fig.~\ref{fig:timeResEnvelopesAndFilter}(b), is made wider than the
95\% confidence level by a factor of two, to allow for the additional
uncertainty associated with the spread of primary vertex locations,
particularly for the endcap calorimeters.  The effect of the primary
vertex distribution is expected in collision data and included in the
CMS simulation but not included in
Fig.~\ref{fig:timeResEnvelopesAndFilter}(b), which is intended to
replicate the results of test beam data. The larger envelope is also
considered a conservative choice for startup when the detector is
expected to be less well calibrated, and other sources of irreducible
uncertainty will not yet have been identified.  For hits with energies
below 4\GeV, no requirement is applied except that the peak amplitude
should be in the triggered data sample.

\begin{figure}[htbp]
  \begin{center}
   \resizebox{\linewidth}{!}{\includegraphics{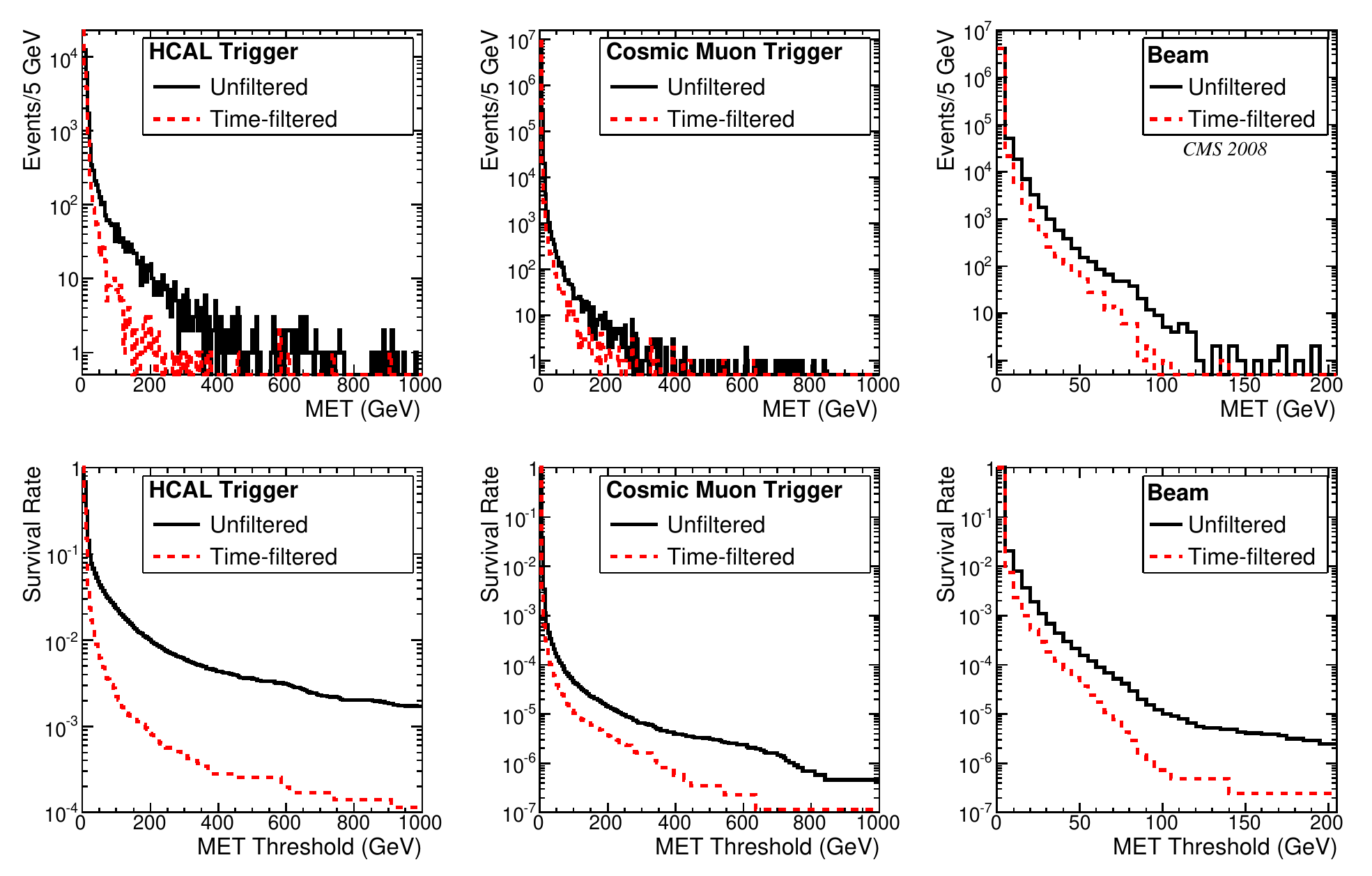}}
    \caption{Time filter performance on MET for HCAL-triggered noise
    events, events selected by the cosmic ray muon trigger, and events
    from beam operations.  The top plots show the MET distributions
    calculated twice for each event, once using all measurements
    (unfiltered) and once using only in-time measurements
    (time-filtered).  The bottom plots show the fraction of events
    that survive as a function of a MET threshold applied
    to both unfiltered and time-filtered measurements.}
    \label{fig:bkgdMETplots}
  \end{center}
\end{figure}

Using the strategy outlined above, the timing filter was then applied
to two datasets: beam background data collected during LHC beam
commissioning in September 2008 and cosmic ray muon and CMS detector
noise background data collected in October and November 2008, during
the month-long data-taking exercise known as the Cosmic Run At Four
Tesla (CRAFT)~\cite{cite:CRAFTGeneral}.  For each event, the MET was
calculated twice for comparison.  One calculation of MET used all
energy deposits (referred to as ``unfiltered''), while the second did
not include energy deposits with a measured time outside the window
(referred to as ``time-filtered'').  The unfiltered measurements
include energy depositions in bunch crossings other than the triggered
one, as a consequence of the triggering and reconstruction algorithms
used during these periods.

The comparison between the unfiltered MET and the time-filtered MET
for HCAL-triggered events, cosmic ray events, and beam events is shown
in Fig.~\ref{fig:bkgdMETplots}. The HCAL-triggered events were
collected using a Level-1 trigger in which a single calorimetric tower
energy exceeds a 5\GeV\ threshold.  Therefore, this sample includes
photo-detector and other instrumental noise
events~\cite{cite:anomevents}.  Events with a Level-1 muon trigger
were removed from the sample.  The cosmic ray muon events were
collected during CRAFT with a Level-1 trigger in which a single muon
is identified by the CMS muon stations. The beam events were taken
with a single circulating 450\GeV\ proton bunch on September 18, 2008.
All events from this sample are included in the plots regardless of
trigger source, which could be beam pickup, coincidence of
scintillator or forward calorimeter energy deposits, or any of several
muon triggers.  The beam sample thus includes beam halo and beam-gas
events as well as cosmic ray muons and accidental overlaps with
calorimeter noise events.  In all cases, the physical MET value would
be zero, so a net migration of events from high MET to low MET is
expected when comparing the unfiltered and time-filtered MET
distributions.

Figure~\ref{fig:bkgdMETplots} also shows the fraction of events that
survive a minimum MET criterion as a function of the threshold
value. This was done for both the time-filtered and unfiltered
distributions for each background type. Figure~\ref{fig:bkgdMETplots},
therefore, indicates that a reduction by a factor of 5--10 in the rate
of events with fake MET can be achieved with time filtering for all
three sources of background studied.  This rejection factor is
consistent with the ratio of the filter window width to the width of
the four time samples considered in the reconstruction of an HCAL
measurement. The fake MET production for all three samples is due
predominantly to a small number of high energy measurements, against
which the filter is most effective.

The filter can be tuned to enhance further the background rejection
performance. More sophisticated algorithms could be employed to filter
clusters of cells with mixed in-time and out-of-time signals if the
members of the cluster share common characteristics that identify them
as background. For instance, beam halo muons may create a cluster of
deposits in consecutive cells along $\eta$ for the same $\phi$, most
of which will be out-of-time. This pattern, once identified, could
allow the removal of nominally in-time background as well.

\subsection{Impact on physics processes}

To estimate the impact of this technique on signals that have not yet
been produced in the CMS detector, studies have been performed on two
simulated processes, one that contains MET at the event generator
level and a second that does not. An interesting example of the first
case is the real MET induced by the hypothetical weakly-interacting
supersymmetric (SUSY)~\cite{cite:susy} particle called the neutralino.
By applying an ill-chosen time filter, one may underestimate the MET
induced by the neutralino by filtering out calorimetric deposits
associated with the recoiling jet.  A typical process without
significant MET at generator level is a hadronic dijet event, commonly
referred to as a QCD dijet event. In this case one may also infer a
(fake) MET in an event in which no MET is expected, causing a tail in
the high end of the MET distribution.  Therefore, an ill-chosen time
filter could degrade the discovery potential of a neutralino search by
reducing the measured MET of the neutralino signal while simultanously
increasing the (fake) MET of a major background.

\begin{figure}[tp]
  \begin{center}
    \resizebox{0.8\linewidth}{!}{\includegraphics{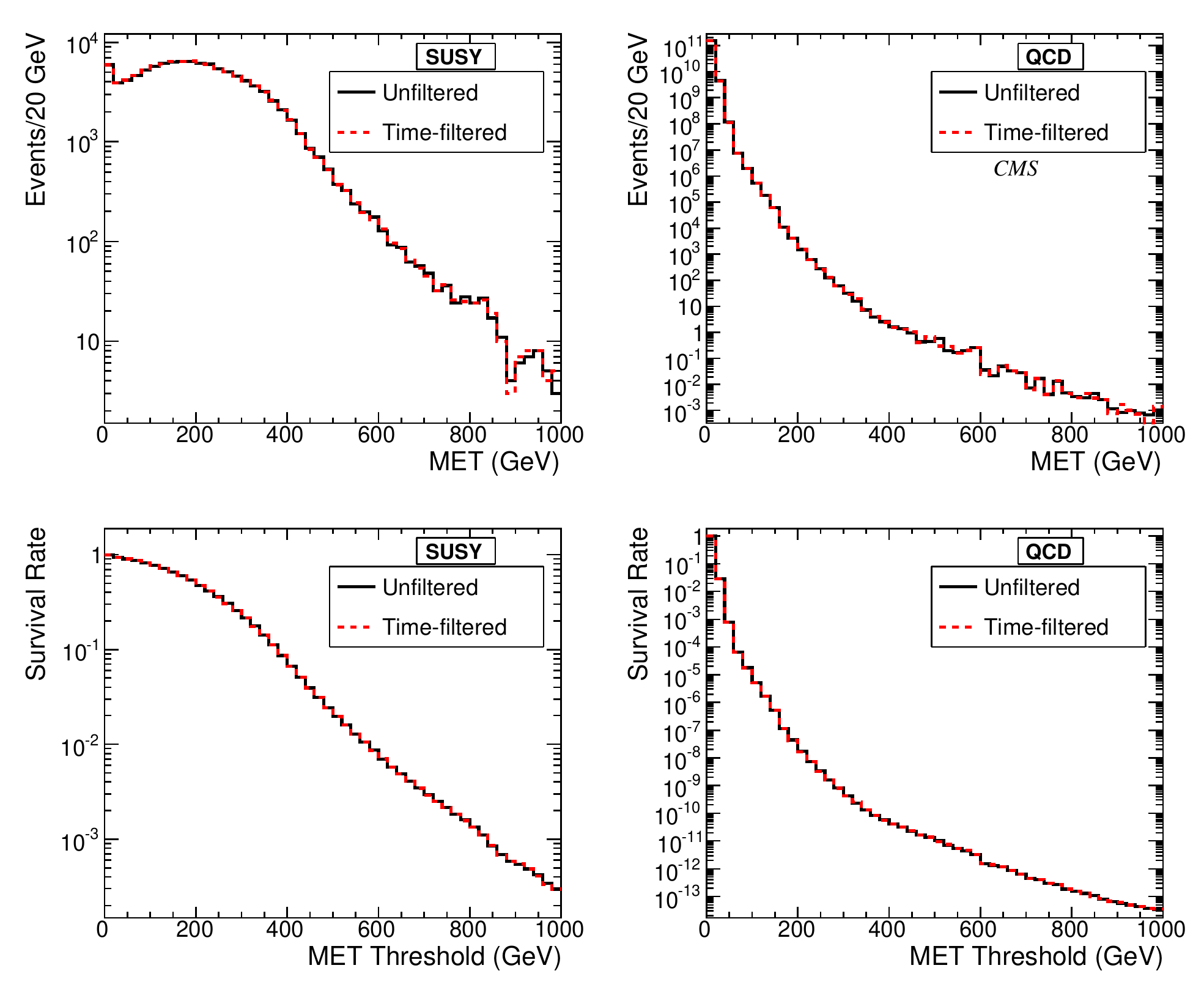}}

    \caption{Time filter performance for SUSY and QCD processes
    showing the spectrum of filtered and unfiltered MET (top plots)
    and the MET survival rates (bottom plots). These indicate that the
    MET is not significantly affected by the operation of the time
    filter for in-time processes. These simulated samples were
    generated at $\sqrt{s}=10$\TeV,
    $L=10^{30}~\mathrm{cm}^{-2}\mathrm{s}^{-1}$, and with no pileup.}

    \label{fig:signalMETplots}
  \end{center}
\end{figure}

To assess whether the chosen timing cuts induce a bias on processes
with significant MET, the timing filter was applied to a sample of
simulated events containing neutralinos and generated under the
assumptions of $\sqrt{s}=10\TeV$ and no overlapping events
($L=10^{30}~\mathrm{cm}^{-2}\mathrm{s}^{-1}$).  The selected SUSY
sample was generated using ISASUSY~\cite{cite:isasusy} and Pythia
6.4~\cite{cite:pythia} at a ``low mass'' point referred to as
``LM1'' \cite{cite:susylm1} in the phase space of one favored model of
SUSY known as mSUGRA \cite{cite:susy}.  The early discovery or
exclusion of this point is potentially within reach of the LHC
experiments given the integrated luminosity expected from the first
physics run.  The left plots in Fig.~\ref{fig:signalMETplots} show
that the same energy-dependent timing window used in
Section \ref{sec:bkgdtiming} preserves MET distributions for simulated
SUSY events with high efficiency.

The impact of the timing filter for QCD dijet events was also studied.
The QCD dijet process was generated by Pythia for a range of hard
scattering scales $15\GeVc < \hat{p}_T < 3\TeVc$ with same
center-of-mass and luminosity assumptions as the SUSY sample.  While
these events have no inherent MET, mismeasurements and dead material
will result in the reconstruction of some MET in these events,
particularly for events with large total transverse energy.  The
results for the QCD dijet sample are shown in the right column of
Fig.~\ref{fig:signalMETplots} and indicate that the MET distribution
is maintained by the filtering process.  It is important to note that
this filter is not intended to remove QCD backgrounds, which are
inherently in-time.  The point of the comparison is to verify that an
increase in MET is not inadvertently created by the filtering process.

In order to apply the time filter to analysis of physics processes,
actual performance of the filter will have to be rederived on
collision data using processes such as $W\rightarrow\ell\nu$ events
containing jets.  The generation of fake MET will be measured using
processes such as $\gamma$~+~jet, where no MET is expected and the
recoil against the photon provides good control of possible jet energy
mismeasurement. The efficiency of background rejection can be further
studied using dijet event samples containing a cosmic ray or beam halo
muon as identified by the muon system.

\section{Summary}

This paper has presented the technique used to determine the arrival
time of energy deposits in the CMS HCAL to less than 25ns, and the use
of this technique to suppress out-of-time backgrounds.  The
performance of the technique and of the synchronization system of the
HCAL have been demonstrated using data from test beam operations as
well as "beam splash" data from the LHC in September 2008.  The
ultimate resolution of the technique for hadron showers was measured
to be 1.2~ns using test beam data, and the time spread of HCAL
channels after hardware alignment in $\eta$ was measured to be
$\pm$2~ns, with an expected potential offline synchronization of less
than 1~ns.  Using data from CRAFT and beam operations in 2008, the use
of a timing filter to suppress fake transverse missing energy was
studied.  This technique was effective in reducing the rate of MET
produced by cosmic ray muons, detector noise, and beam background by a
factor of 5--10.  At the same time, the integrity of simulated signal
MET distributions under the same filtering technique was verified.
This technique and the timing results from HCAL in general are
available for use in CMS analyses to suppress backgrounds or select
particular signals.

\section*{Acknowledgments}

We thank the technical and administrative staff at CERN and other CMS Institutes,
and acknowledge support from:
FMSR (Austria);
FNRS and FWO (Belgium);
CNPq, CAPES, FAPERJ, and FAPESP (Brazil);
MES (Bulgaria);
CERN;
CAS, MoST, and NSFC (China);
COLCIENCIAS (Colombia);
MSES (Croatia);
RPF (Cyprus);
Academy of Sciences and NICPB (Estonia);
Academy of Finland, ME, and HIP (Finland);
CEA and CNRS/IN2P3 (France);
BMBF, DFG, and HGF (Germany);
GSRT (Greece);
OTKA and NKTH (Hungary);
DAE and DST (India);
IPM (Iran);
SFI (Ireland);
INFN (Italy);
NRF (Korea);
LAS (Lithuania);
CINVESTAV, CONACYT, SEP, and UASLP-FAI (Mexico);
PAEC (Pakistan);
SCSR (Poland);
FCT (Portugal);
JINR (Armenia, Belarus, Georgia, Ukraine, Uzbekistan);
MST and MAE (Russia);
MSTDS (Serbia);
MICINN and CPAN (Spain);
Swiss Funding Agencies (Switzerland);
NSC (Taipei);
TUBITAK and TAEK (Turkey);
STFC (United Kingdom);
DOE and NSF (USA).
Individuals have received support from
the Marie-Curie IEF program (European Union);
the Leventis Foundation;
the A. P. Sloan Foundation;
and the Alexander von Humboldt Foundation.

\bibliography{auto_generated}   

\providecommand{\href}[2]{#2}\begingroup\raggedright\begin{thebibliography}{10}

\bibitem{cite:jinstcms}
{CMS Collaboration}, ``The CMS Experiment at the CERN LHC'', {\em J. Inst.}
  {\bf {3 S08004}} (2008).
  \href{http://dx.doi.org/10.1088/1748-0221/3/08/S08004}{{\tt
  doi:10.1088/1748-0221/3/08/S08004}}.

\bibitem{cite:jinstlhc}
{L. Evans and P. Bryant (eds.)}, ``LHC Machine'', {\em J. Inst.} {\bf {3
  S08001}} (2008). \href{http://dx.doi.org/10.1088/1748-0221/3/08/S08001}{{\tt
  doi:10.1088/1748-0221/3/08/S08001}}.

\bibitem{cite:cmsHSCPnote}
{CMS Collaboration}, ``Search for Heavy Stable Charged Particles'', {\em CMS
  PAS} {\bf
  \href{http://cms-physics.web.cern.ch/cms-physics/public/EXO-08-003-pas.pdf}{%
EXO-08-003}} (2008).

\bibitem{cite:epjcHB}
{CMS-HCAL Collaboration}, ``Design, performance, and calibration of CMS
  hadron-barrel calorimeter wedges'', {\em Eur. Phys. J. C} {\bf 55} (2008)
  {159--171}. \href{http://dx.doi.org/10.1140/epjc/s10052-008-0573-y}{{\tt
  doi:10.1140/epjc/s10052-008-0573-y}}.

\bibitem{cite:jinstcmsHE}
{CMS Collaboration}, ``The CMS Experiment at the CERN LHC'', {\em J. Inst.}
  {\bf {3 S08004}} (2008) {131--137}.
  \href{http://dx.doi.org/10.1088/1748-0221/3/08/S08004}{{\tt
  doi:10.1088/1748-0221/3/08/S08004}}.

\bibitem{cite:epjcHO}
{CMS-HCAL Collaboration}, ``Design, performance, and calibration of the CMS
  hadron-outer calorimeter'', {\em Eur. Phys. J. C} {\bf 57} (2008) {653--663}.
  \href{http://dx.doi.org/10.1140/epjc/s10052-008-0756-6}{{\tt
  doi:10.1140/epjc/s10052-008-0756-6}}.

\bibitem{cite:qie}
T.~Zimmerman and J.~Hoff, ``The Design of a Charge Integrating, Modified
  Floating Point ADC Chip'', {\em IEEE J. Solid State Circuits} {\bf {39}}
  (2004) {895--905}. \href{http://dx.doi.org/10.1109/JSSC.2004.827808}{{\tt
  doi:10.1109/JSSC.2004.827808}}.

\bibitem{cite:craftL1trig}
{CMS Collaboration}, ``Performance of the CMS Level-1 Trigger during
  Commissioning with Cosmic Rays'', {\em submitted to JINST} (2009).

\bibitem{cite:timeslew}
{CMS Collaboration}, ``The CMS Experiment at the CERN LHC'', {\em J. Inst.}
  {\bf {3 S08004}} (2008) {152--154}.
  \href{http://dx.doi.org/10.1088/1748-0221/3/08/S08004}{{\tt
  doi:10.1088/1748-0221/3/08/S08004}}.

\bibitem{cite:epjcHBTB}
{CMS-HCAL Collaboration}, ``The CMS barrel calorimeter response to particle
  beam from 2 to 350 GeV/c'', {\em Eur. Phys. J. C} {\bf 60} (2009) {359--373}.
  \href{http://dx.doi.org/10.1140/epjc/s10052-009-0959-5}{{\tt
  doi:10.1140/epjc/s10052-009-0959-5}}.

\bibitem{cite:cmsMETnote}
H.~Pi, P.~Avery, D.~Green{ et~al.}, ``Measurement of Missing Transverse Energy
  With the CMS Detector at the LHC'', {\em CMS NOTE} {\bf
  \href{http://cms.cern.ch/iCMS/jsp/openfile.jsp?type=NOTE&year=2006&files=NOT%
E2006_035.pdf}{2006/035}} (2006).

\bibitem{cite:CRAFTGeneral}
{CMS Collaboration}, ``The CMS Cosmic Run at Four Tesla'', {\em submitted to
  JINST} (2009).

\bibitem{cite:anomevents}
{CMS Collaboration}, ``Identification and Filtering of Atypical Signals in the
  CMS Hadronic Calorimeter'', {\em submitted to JINST} (2009).

\bibitem{cite:susy}
S.~P. Martin, ``{A Supersymmetry Primer}'',
\href{http://www.arXiv.org/abs/hep-ph/9709356}{{\tt arXiv:hep-ph/9709356}}.

\bibitem{cite:isasusy}
H.~Baer, F.~E. Paige, S.~D. Protopopescu{ et~al.}, ``{Simulating supersymmetry
  with ISAJET 7.0 / ISASUSY 1.0}'',
\href{http://www.arXiv.org/abs/hep-ph/9305342}{{\tt arXiv:hep-ph/9305342}}.

\bibitem{cite:pythia}
T.~Sjostrand, S.~Mrenna, and P.~Skands, ``{PYTHIA 6.4 Physics and Manual}'',
  {\em JHEP} {\bf 05} (2006)
026.
  \href{http://dx.doi.org/10.1088/1126-6708/2006/05/026}{{\tt
  doi:10.1088/1126-6708/2006/05/026}}.

\bibitem{cite:susylm1}
{CMS Collaboration}, ``CMS Physics Technical Design Report, Volume II: Physics
  Performance'', {\em J. Phys. G: 382 Nucl. Part. Phys.} {\bf 34} (2007) 1387.
  \href{http://dx.doi.org/10.1088/0954-3899/34/6/S01}{{\tt
  doi:10.1088/0954-3899/34/6/S01}}.

\end{thebibliography}\endgroup
\clearpage
\cleardoublepage\appendix\section{The CMS Collaboration \label{app:collab}}\begin{sloppypar}\hyphenpenalty=500\textbf{Yerevan Physics Institute,  Yerevan,  Armenia}\\*[0pt]
S.~Chatrchyan, V.~Khachatryan, A.M.~Sirunyan
\vskip\cmsinstskip
\textbf{Institut f\"{u}r Hochenergiephysik der OeAW,  Wien,  Austria}\\*[0pt]
W.~Adam, B.~Arnold, H.~Bergauer, T.~Bergauer, M.~Dragicevic, M.~Eichberger, J.~Er\"{o}
% ...
\vskip\cmsinstskip
% ...

\end{sloppypar}
\end{document}